\documentclass[aps,prx,twocolumn,showpacs]{revtex4-1}
\usepackage{bm}
\usepackage{mathrsfs}
\usepackage{amsmath}
\usepackage{amssymb}
\usepackage{graphicx}
\usepackage{amsfonts}
\usepackage{amsthm}
\usepackage{color}
\usepackage{dcolumn}
\usepackage{txfonts}
\usepackage{subfig}
\newcommand{\RN}[1]{%
  \textup{\uppercase\expandafter{\romannumeral#1}}%
}

\begin{document}

\title{Analytical Approximations to the Dynamics of Nonlinear Level Crossing Models}
\author{Chon-Fai Kam}
\author{Yang Chen}
\email{Corresponding author. Email: yangbrookchen@yahoo.co.uk}
\affiliation{Department of Mathematics, Faculty of Science and Technology,
University of Macau, Avenida da Universidade, Taipa, Macau, China}

\begin{abstract}
We study the dynamics of a nonlinear two-level crossing model with a cubic modification of the linear Landau-Zener diabatic energies. The solutions are expressed in terms of the bi-confluent Heun functions --- the generalization of the confluent hypergeometric functions. We express the finial transition probability as a convergent series of the parameters of the nonlinear laser detuning, and derive analytical approximations for the state populations in terms of parabolic cylinder and Whittaker functions. Tractable closed-form expressions are derived for a large part of the parameter space. We also provide simple method to determine the transition point which connects local solutions in different physical limits. The validity of the analytical approximations is shown by comparison with numerical results of simulations. 
\end{abstract}

\maketitle

\section{indroduction}
Level crossing models \cite{nakamura2012nonadiabatic} which are crucial for understanding non-adiabatic transitions in physics, chemistry and biology, have a long and interesting history in science, and have recently regained popularity in areas such as non-Hermitian systems \cite{longhi2017oscillating, gong2018piecewise, zhang2018dynamically, zhang2018hybrid, longstaff2019non}, and quantum molecular dynamics \cite{curchod2018ab, crespo2018recent, daligault2018nonadiabatic, zhang2019non}. It was first formulated by Landau \cite{landau1932theoryI, landau1932theory} in 1932 for analyzing atomic collisions in both sudden and adiabatic limits, and was analytically solved by Zener using parabolic cylinder functions \cite{zener1932non}. Subsequently, St{\"u}ckelberg derived a sophisticated tunneling formula based on analytical continuation of the semi-classical WKB solutions across the Stokes lines \cite{stuckelberg1932theory}, and Majorana derived the transition probability formula using integral representation of the survival amplitude in the context of the dynamics of a spin-1/2 in a time-dependent magnetic field \cite{majorana1932atomi}. The Landau-Zener model assumes a constant coupling between bare states in the diabatic basis, and a linearly varying separation of diabatic energies \cite{wittig2005landau}. The tunneling probability $e^{-\kappa}$ in the Landau-Zener model only depends on a single dimensionless parameter $\kappa\equiv 2\pi f^2/(\hbar|(F_1-F_2)V|)$, where $f$ is the coupling matrix element in the diabatic basis, $F_1$ and $F_2$ are the slopes of the intersecting diabatic potential curves, and $V$ is the velocity of the perturbation variable, \textit{e.g.}, the relative collision velocity \cite{di2005majorana}. Since Landau, Zener, St{\"u}ckelberg, and Majorana's pioneering works in the 1930s, the linear two-state model has been applied to vast areas like atomic \cite{smirnov2003physics, nikitin2012theory} and molecular collisions \cite{child2010molecular}, atoms in intense laser fields \cite{delone1985atoms, kazantsev1990mechanical}, spin tunneling in molecular nano-magnets \cite{wernsdorfer1999quantum}, tunneling of Bose-Einstein condensates in accelerated optical lattices \cite{morsch2001bloch}, and optical tunneling in waveguide arrays \cite{khomeriki2005nonadiabatic}.

Although the Landau-Zener model has achieved great success over the last century, there are indeed cases where the assumption of linear crossing between the diabatic states breaks down. Indeed, one may employ the St{\"u}ckelberg tunneling formula to the individual tunneling events \cite{stuckelberg1932theory, shevchenko2010landau}, as long as any pair of Landau-Zener crossings are well-separated from each other. But for cases where the crossing points merge together as a result of external parameter changes, the Landau-Zener linearization fails, and the linear-dependence of diabatic energies should be replaced by a superlinear or sublinear one \cite{garraway1995wave}. The simplest nonlinear level crossing model is the parabolic model with a constant coupling, and a quadratically varying detuning, which was first introduced by Bikhovskii \textit{et. al.} in 1965 in the context of slow atomic collisions \cite{bikhovskii1965probability} and was later re-evaluated by Delos and Thorson \cite{delos1972solution, delos1972studies} and Crothers \cite{crothers1975stueckelberg, crothers1976perturbed, crothers1977stueckelberg} in various physical limits. In the 1990s, Shimshoni and Gefen incorporated environment-induced dissipation and dephasing into the parabolic model \cite{shimshoni1991onset}, and Suominen derived analytical approximations for the final state populations \cite{suominen1992parabolic}, and applied the results to the dynamics of cold atoms in magnetic traps \cite{pakarinen2000atomic}. In the same period, Zhu and Nakamura derived exact formulas for the scattering matrices in terms of a convergent infinite series \cite{zhu1992two, zhu1992twoII, zhu1993two, zhu1994two, zhu2001nonadiabatic}. Nakamura \textit{et. al.} applied the results to laser assisted surface ion neutralization \cite{teranishi1997semiclassical} and laser-controlled photochromism in functional molecules \cite{tamura2006laser}. Over the last decade, Lehto incorporated super-parabolic level glancing effects into the parabolic model \cite{lehto2012superparabolic, lehto2013zhu}, and revealed complete population inversion due to phase-jump couplings \cite{lehto2016two}. Zhang \textit{et. al.} expressed the population dynamics of driven dipolar molecules in the parabolic level glancing model in terms of the confluent Heun functions \cite{zhang2016analytic}. Recently, the parabolic model has found applications in topological systems such as interband tunneling of fermionic atoms near the merging transition of Dirac cones in tunable honeycomb optical lattices \cite{fuchs2012interband}, or interband tunneling of two-dimensional electrons near the topological transition in type II Weyl semimetals \cite{malla2019high}.

The tunneling dynamics of parabolic level crossing models may be described by the tri-confluent Heun function \cite{ronveaux1995heun}, which is obtained from the general Heun function \cite{heun1888theorie} by coalescencing three regular singularities with infinity. Unlike the parabolic cylinder function used in the Landau-Zener model, there are no comprehensive information about the asymptotic behaviors of Heun functions because of the lack of integral representations in terms of simpler functions. In other words, Majorana's integral representation method cannot be used to determine the final transition probability. Nevertheless, in a recent work \cite{kam2019analytical}, we showed that analytical approximations to the transition amplitudes via simple functions can still be obtained in a large part of the parameter space. In this work, we studied a type of level crossing models where the coupling matrix element is a constant, and the nonlinear detuning is an odd function of time. In particular, we derived analytical approximations to the transition amplitudes for the case when the linear Landau-Zener detuning is modified by a cubic nonlinearity in the vicinity of the crossing. The nonlinear level crossing model was first revealed by Vitanov and Suominen by using the Dykhne-Davis-Pechukas formula \cite{vitanov1999nonlinear}, and was subsequently found applications in areas such as interaction assisted tunneling in Bose-Einstein condensates \cite{zobay2000time}, formation of ultracold molecules by photoassociatioin process \cite{vala2000coherent}, and high-fidelity adiabatic quantum computation \cite{barankov2008optimal, torosov2011high}.

The paper is organized as follows: In Section \ref{II}, we introduce the nonlinear level crossing model in the context of a two-level atom driven by a classical laser field, in which the linear Landau-Zener detuning is modified by a cubic nonlinearity in the vicinity of the crossing, and the Rabi frequency for laser driven atomic transitions remains time-independent. In Section \ref{III}, we express the final state population in an analytical form as a convergent infinite series of the parameters of the nonlinear laser detuning and the Rabi frequency. In Section \ref{IV}, we derive tractable analytical approximations for the transition amplitudes in both short- and long-time limits. The accuracy of the analytical approximations is tested via comparison with results obtained from numerical simulations. We determine the transition point by minimizing the maximal error caused by using simple special functions in different physical limits. Finally, in Section \ref{V}, we conclude our study, and outline possible extensions and applications of our work.

\section{The nonlinear level crossing model}\label{II}
Within the rotating-wave approximation, the wave amplitudes of a two-level atom dipole-interacting with a classical electric field are governed by the following coupled equations
\begin{equation}
i\frac{da_1}{dt}=-\frac{\Delta}{2}a_1+fa_2,\:
i\frac{da_2}{dt}=f^*a_1+\frac{\Delta}{2}a_2,
\end{equation}
where $\Delta$ is the laser detuning, and $f$ is the Rabi frequency at resonance. After applying the transformations $a_1=C_1e^{\frac{i}{2}\int_0^t\Delta ds}$ and $a_2=C_2e^{-\frac{i}{2}\int_0^t\Delta ds}$, the coupled equations becomes
\begin{equation}
    i\frac{dC_1}{dt}=fe^{-i\int_0^t\Delta ds}C_2,\:
    i\frac{dC_2}{dt}=f^*e^{i\int_0^t\Delta ds}C_1,
\end{equation}
or equivalently
\begin{subequations}
\begin{gather}
    \frac{d^2C_1}{dt^2}+\left(i\Delta-\frac{\dot{f}}{f}\right)\frac{dC_1}{dt}+|f|^2C_1=0,\label{TwoLevelProblem2}\\
    \frac{d^2C_2}{dt^2}-\left(i\Delta+\frac{\dot{f}^*}{f^*}\right)\frac{dC_2}{dt}+|f|^2C_2=0,\label{TwoLevelProblem1}
\end{gather}
\end{subequations}
where $C_1$ and $C_2$ satisfy the normalization condition $|C_1|^2+|C_2|^2=1$. The problem of non-adiabatic transition is to determine the transition probability $|C_1(\infty)|^2$, subjected to the conditions $|C_1(-\infty)|=0$ and  $|C_2(-\infty)|=1$. Using the change of variable $C_1=U_1\exp\{-\frac{1}{2}\int_0^tp ds\}$, where $p\equiv i\Delta-\dot{f}/f$, we obtain the Schr\"{o}dinger form of Eq.\:\eqref{TwoLevelProblem2}
\begin{subequations}
\begin{gather}
    \frac{d^2U_1}{dt^2}+J(t)U_1=0,\\
    J(t)=|f|^2-\frac{1}{2}\frac{d}{dt}\left(i\Delta-\frac{\dot{f}}{f}\right)-\frac{1}{4}\left(i\Delta-\frac{\dot{f}}{f}\right)^2.
\end{gather}
\end{subequations}

In the previous work \cite{kam2019analytical}, we analyzed the non-adiabatic dynamics of the parabolic model, in which the laser detuning is a quadratic function of time, $\Delta(t)=\alpha t+\beta t^2$. In the current work, we will analyze the another case, in which the laser detuning is a cubic function of time, $\Delta(t)=\alpha t+\gamma t^3$, and the Rabi frequency $f$ remains a time-independent constant. The equation which governs the wave amplitude $C_1$, subjected to the conditions $|C_1(-\infty)|=0$ and $|C_1(\infty)|=1$, becomes
\begin{equation}\label{SuperParabolic}
   \ddot{C}_1+i\left(\alpha t+\gamma t^3\right)\dot{C}_1+|f|^2C_1=0.
\end{equation}
Then the differential equation for $U_1\equiv C_1\exp\{\frac{i}{2}\int_0^t\Delta(s)ds\}$ becomes
\begin{equation}
    \ddot{U}_1+\left[|f|^2-\frac{i\alpha}{2}+\left(\frac{\alpha^2}{4}-\frac{3i\gamma}{2}\right)t^2+\frac{\alpha\gamma}{2} t^4+\frac{\gamma^2}{4}t^6\right]U_1=0,
\end{equation}
which is equivalent to the Schr\"{o}dinger equation for a doubly anharmonic oscillator $\psi^{\prime\prime}+(E-(\mu x^2+\lambda x^4+\eta x^6))\psi=0$ with $E=|f|^2-\frac{\alpha}{2}i$, $\mu=-\frac{1}{4}\alpha^2+\frac{3}{2}\gamma i$, $\lambda=-\frac{1}{2}\alpha\gamma$ and $\eta=-\frac{1}{4}\gamma^2$. After the change of variable $z=t^2$, Eq.\:\eqref{SuperParabolic} becomes
\begin{equation}\label{BiConfluentHeun}
    C_1^{\prime\prime}+\frac{1}{2}\left(\frac{1}{z}+i\alpha+i\gamma z\right)C_1^\prime+\frac{|f|^2}{4z}C_1=0.
\end{equation}
After a change of variable $W_1\equiv C_1z^{\frac{1}{4}}\exp\{\frac{i}{4}(\alpha z+\frac{\gamma}{2}z^2)\}$, we obtain
\begin{equation}\label{SchrodingerEquationWt}
    W_1^{\prime\prime}=-\left[\frac{3}{16z^2}+(\frac{|f|^2}{4}-\frac{i\alpha}{8})\frac{1}{z}+(\frac{\alpha^2}{16}-\frac{3i\gamma}{8})+\frac{\alpha\gamma z}{8}+\frac{\gamma^2z^2}{16}\right]W_1,
\end{equation}
which is equivalent to the Schr\"{o}dinger equation for a potential $V=ax^{-2}+bx^{-1}+cx+dx^2$ in the form $\psi^{\prime\prime}+(E-V(x))\psi=0$, where $E=\frac{1}{16}(\alpha^2-6i\gamma)$, $a=-\frac{3}{16}$, $b=-\frac{1}{4}|f|^2+\frac{1}{8}i\alpha$, $c=-\frac{1}{8}\alpha\gamma$ and $d=-\frac{1}{16}\gamma^2$. After a further change of variable $w=\sqrt{-i\gamma}z/2$, Eq.\:\eqref{BiConfluentHeun} becomes the canonical form of the bi-confluent Heun equation \cite{ronveaux1995heun, heun1888theorie}
\begin{equation}
    C_1^{\prime\prime}+\left[\frac{1+\mu}{w}-\nu-2w\right]C_1^\prime+\left[\lambda-\mu-2-\frac{\eta+(1+\mu)\nu}{2w}\right]C_1=0,
\end{equation}
or equivalently
\begin{equation}
    W_1^{\prime\prime}+\left(\frac{1-\mu^2}{4w^2}-\frac{\eta}{2w}+\lambda-\frac{\nu^2}{4}-\nu w-w^2\right)W_1=0,
\end{equation}
where
\begin{equation}
    \mu=-\frac{1}{2},\nu=\frac{\alpha\sqrt{-i\gamma}}{\gamma},\lambda=\mu+2,\eta=-\frac{\nu}{2}-\frac{|f|^2}{\sqrt{-i\gamma}}.
\end{equation}

\section{Explicit Formulas for the transition amplitudes at infinity}\label{III}
\begin{figure}[tbp]
\begin{center}
\includegraphics[width=0.6\columnwidth]{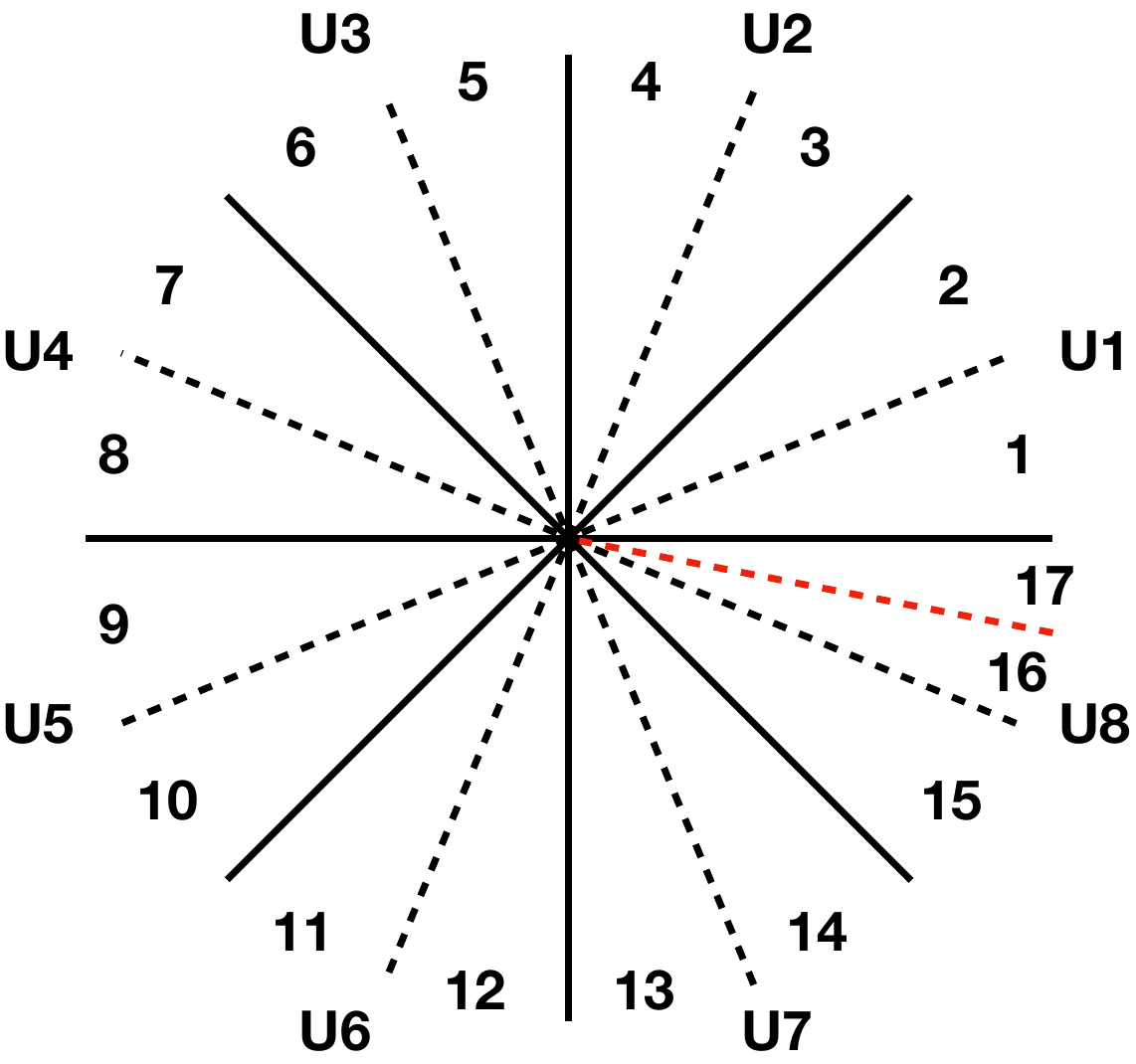}
\end{center}
\caption{Stokes and anti-Stokes lines on the complex $t$ plane, which are labeled by dashed and solid lines respectively. The branch cut are labeled by red dashed line. There are in total eight Stokes constants.}
\label{fig:WKBDiagram}
\end{figure}

As we have shown in the last section, for a nonlinear level crossing model, in which the linear Landau-Zener detuning is modified by a cubic non-linearity in the vicinity of the crossing, $\Delta(t)=\alpha t+\gamma t^3$, the equation which governs the wave amplitude $U_1(t)$ has the form
\begin{equation}\label{BCHeun}
    \ddot{U}_1=-(a_6t^6+a_4t^4+a_2t^2+a_0)U_1\equiv -q(t)U_1,
\end{equation}
where $a_6=\frac{1}{4}\gamma^2$, $a_4=\frac{1}{2}\alpha\gamma$, $a_2=\frac{1}{4}\alpha^2-\frac{3}{2}i\gamma$ and $a_0=|f|^2-\frac{1}{2}i\alpha$. In this section, we discuss the solutions of Eq.\:\eqref{BCHeun} around the irregular singularity $t=\infty$. Note that when $|t|$ is large, one has the asymptotic relation $q(t)\rightarrow q_6t^6$. Hence, the asymptotic WKB solutions of $U_1(t)$ behave as $t^{-3/2}\exp{(\pm\frac{i}{4}\sqrt{a_6}t^4)}$. The eight anti-Stokes lines are then asymptotically defined by $\Re(it^4)=0$, or $\arg{t}=k\pi/4$; whereas the eight Stokes lines are asymptotically defined by $\Im(it^4)=0$, or $\arg{t}=k\pi/4+\pi/8$ (see Fig.\:\ref{fig:WKBDiagram}). Following Heading's notation \cite{heading2013introduction}, we use the symbol $(\bullet,t)$ to denote the WKB solution $q^{-1/4}\exp(i\int_\bullet^t \sqrt{q(t^\prime)}dt^\prime)$, where the notation $\bullet$ means that the lower limit of integration is not specified. Similarly, $(t,\bullet)\equiv q^{-1/4}\exp(-i\int_\bullet^t \sqrt{q(t^\prime)}dt^\prime)$, where the minus sign reverses the path of integration. Hence, the WKB solutions near the irregular singularity $t=\infty$ are
\begin{subequations}
\begin{align}
    (\bullet,t)&=t^{-\frac{3}{2}}\exp\left\{i\left[\frac{\sqrt{a_6}}{4}t^4+\frac{a_4}{4\sqrt{a_6}}t^2+\frac{1}{2\sqrt{a}_6}(a_2-\frac{a_4^2}{4a_6})\ln t\right]\right\}\nonumber\\
    &=t^{-\frac{3}{2}}\exp\left\{\pm i\left(\frac{\gamma}{8}t^4+\frac{\alpha}{4}t^2-\frac{3i}{2}\ln t\right)\right\},\\
    (t,\bullet)&=t^{-\frac{3}{2}}\exp\left\{-i\left[\frac{\sqrt{a_6}}{4}t^4+\frac{a_4}{4\sqrt{a_6}}t^2+\frac{1}{2\sqrt{a}_6}(a_2-\frac{a_4^2}{4a_6})\ln t\right]\right\}\nonumber\\
    &=t^{-\frac{3}{2}}\exp\left\{\mp i\left(\frac{\gamma}{8}t^4+\frac{\alpha}{4}t^2-\frac{3i}{2}\ln t\right)\right\},
\end{align}
\end{subequations}
where the plus and minus signs are for $\gamma>0$ and $\gamma<0$ respectively. For $\gamma>0$, we have $(\bullet,t)=e^{i(\frac{\gamma}{8}t^4+\frac{\alpha}{4}t^2)}$ and $(t,\bullet)=t^{-3}e^{-i(\frac{\gamma}{8}t^4+\frac{\alpha}{4}t^2)}$; whereas for $\gamma<0$, we have $(\bullet,t)=t^{-3}e^{i(\frac{|\gamma|}{8}t^4-\frac{\alpha}{4}t^2)}$ and $(t,\bullet)=e^{-i(\frac{|\gamma|}{8}t^4-\frac{\alpha}{4}t^2)}$. We now perform the transformations $\tau\equiv \frac{i}{2}\sqrt{a_6}t^4$ and
$T_1=t^{3/2}U_1$, and obtain
\begin{gather}
\frac{d^2T_1}{d\tau^2}=(\frac{1}{4}+\sum_{n=1}^4Q_n\tau^{-\frac{n}{2}})T_1\:\:\mbox{with}\:\:
Q_1\equiv\frac{\pm\alpha e^{\frac{i\pi}{4}}}{4\sqrt{|\gamma|}},\nonumber\\
Q_2\equiv\frac{i(\frac{1}{4}\alpha^2-\frac{3}{2}i\gamma)}{4|\gamma|},
Q_3\equiv\frac{(|f|^2-\frac{i\alpha}{2})e^{\frac{3i\pi}{4}}}{8\sqrt{|\gamma|}},
Q_4\equiv-\frac{15}{64}.\label{Qn}
\end{gather}
We then obtain the following two independent solutions
\begin{equation}\label{IndependentUandV}
    u(\tau;Q_n)\equiv(\bullet,\tau)\sum_{n=0}^\infty c_n\tau^{-\frac{n}{2}},\:\:
    v(\tau;Q_n)\equiv(\tau,\bullet)\sum_{n=0}^\infty d_n\tau^{-\frac{n}{2}},
\end{equation}
where $(\bullet,\tau)=(\tau,\bullet)^{-1}\equiv \tau^\rho e^{\frac{1}{2}\tau 2Q_1\tau^{\frac{1}{2}}}$, $\rho\equiv Q_2-Q_1^2=\pm\frac{3}{8}$ and $d_n\equiv i^nc_n$. The coefficients $c_n$ satisfy the following recurrence relations
\begin{align}
   nc_n&=\left\{4Q_1\left[\rho+\left(\frac{1}{4}-\frac{n}{2}\right)\right]-2Q_3\right\}c_{n-1}\nonumber\\
   &+2\left[\rho^2+\rho(1-n)+\frac{15}{64}+\frac{n}{2}\left(\frac{n}{2}-1\right)\right]c_{n-2},
\end{align}
where $c_0\equiv 1$ and $c_1\equiv 4Q_1(\rho-\frac{1}{4})-2Q_3$. If we denote $\bar{u}\equiv u(e^{-i\pi}\tau;-iQ_1,-Q_2,iQ_3)$ and $\bar{v}\equiv v(e^{-i\pi}\tau;-iQ_1,-Q_2,iQ_3)$, we immediately obtain the relations $\bar{u}=e^{i\pi\rho}v$ and $\bar{v}=e^{-i\pi\rho}u$.

We now discuss the relationships between the Stokes constants on the complex $t$ plane. We use the subscripts $d$ and $s$ to denote the corresponding dominant and sub-dominant asymptotic WKB solutions, \textit{e.g.}, $(t,\bullet)_d$ is dominant in $0<\arg t<\pi/4$. When crossing the anti-Stokes lines $\arg t=k\pi/4$, the property of dominance exchanged, \textit{i.e.}, $(\bullet,t)_s\leftrightarrow(\bullet,t)_d$, $(\bullet,t)_d\leftrightarrow(\bullet,t)_s$. When crossing a branch cut, the solution is multiplied by a phase factor to ensure continuity of the WKB solutions. For a clockwise crossing, $(\bullet,t)\rightarrow -e^{-8\pi i\rho}(\bullet,t)$ and $(t,\bullet)\rightarrow -e^{8\pi i\rho}(t,\bullet)$. The branch cut is located at $\arg t=-\pi/16$ (see Fig.\:\ref{fig:WKBDiagram}). When crossing the Stokes lines $\arg t=k\pi/4+\pi/8$, the sub-dominant solution experiences a jump that is proportional to the Stokes constant, \textit{e.g.}, $A(\bullet,t)_s+B(t,\bullet)_d\rightarrow (A+BU)(\bullet,t)_s+B(t,\bullet)_d$ for a counter-clockwise crossing, where $U$ is the Stokes constant. Within the sector $(k-1)\pi/8<\arg t<k\pi/8$, one may denote the associated WKB solution as $U_1^{(k)}(t)$. Then, for a counter-clockwise crossing, from $U_1^{(1)}(t)=A(\bullet,t)_s+B(t,\bullet)_d$, one obtains
\begin{subequations}
\begin{align}
    U_1^{(2)}(t)&=(A+BU_1)(\bullet,t)_s+B(t,\bullet)_d,\\
    U_1^{(4)}(t)&=(A+BU_1)(\bullet,t)_d\nonumber\\
    &+[AU_2+B(1+U_1U_2)](t,\bullet)_s,\\
    U_1^{(6)}(t)&=[A(1+U_2U_3)+B(U_1+U_3+U_1U_2U_3)](\bullet,t)_s\nonumber\\
    &+[AU_2+B(1+U_1U_2)](t,\bullet)_d,\\
    U_1^{(8)}(t)&=[A(1+U_2U_3)+B(U_1+U_3+U_1U_2U_3)](\bullet,t)_d\nonumber\\
    &+\{A(U_2+U_4+U_2U_3U_4)+B[1+U_1U_2\nonumber\\
    &+U_4(U_1+U_3+U_1U_2U_3)]\}(t,\bullet)_s,\label{Upper}
\end{align}
\end{subequations}
where $U^{(2k+1)}(t)$ is obtained from $U^{(2k)}(t)$ by $(\bullet,t)_d\leftrightarrow(\bullet,t)_s$ and $(t,\bullet)_d\leftrightarrow(t,\bullet)_s$. For a clock-wise crossing, one obtains $U_1^{(17)}(t)=A(\bullet,t)_d+B(t,\bullet)_s$, $U_1^{(16)}(t)=A'(\bullet,t)_d+B'(t,\bullet)_s$ on the lower-half $t$-plane, and
\begin{subequations}
\begin{align}
    U_1^{(15)}(t)&=A'(0,t)_d+(B'-A'U_8)(t,0)_s,\\
    U_1^{(13)}(t)&=[A'(1+U_7U_8)-B'U_7](0,t)_s\nonumber\\
    &+(B'-A'U_8)(t,0)_d,\\
    U_1^{(11)}(t)&=[A'(1+U_7U_8)-B'U_7](0,t)_d+[B'(1+U_6U_7)\nonumber\\
    &-A'(U_6+U_8+U_6U_7U_8)](t,0)_s,\\
    U_1^{(9)}(t)&=\{A'[1+U_7U_8+U_5(U_6+U_8+U_6U_7U_8)]\nonumber\\
    &-B'(U_5+U_7+U_5U_6U_7)\}(0,t)_s+[B'(1+U_6U_7)\nonumber\\
    &-A'(U_6+U_8+U_6U_7U_8)](t,0)_d.\label{Lower}
\end{align}
\end{subequations}
where $A'\equiv-e^{-8\pi i\rho}A$ and $B'\equiv-e^{8\pi i\rho}B$, and $U^{(2k)}$ is obtained from $U^{(2k+1)}$ by $(\bullet,t)_d\leftrightarrow(\bullet,t)_s$ and $(t,\bullet)_d\leftrightarrow(t,\bullet)_s$. On the anti-Stokes line $\arg t=\pi$, identifying Eqs.\:\eqref{Upper} and \eqref{Lower}, one obtains the following relations among the Stokes constants $U_k$
\begin{align}
    &1+U_2U_3=-[1+U_7U_8+U_5(U_6+U_8+U_6U_7U_8)]e^{-8\pi i\rho},\nonumber\\
    &1+U_6U_7=-[1+U_1U_2+U_4(U_1+U_3+U_1U_2U_3)]e^{-8\pi i\rho},\nonumber\\
    &U_1+U_3+U_1U_2U_3=(U_5+U_7+U_5U_6U_7)e^{8\pi i\rho},\nonumber\\
    &U_2+U_4+U_2U_3U_4=(U_6+U_8+U_6U_7U_8)e^{-8\pi i\rho}.
\end{align}

Next, we examine the relations between the Stoke constants $T_k$ on the complex $\tau$ plane and $U_k$ on the complex $t$ plane. In the sector $0<\arg t<\pi/4$, the solution $U_1(t)=B(t,\bullet)$ yields $U_1(t)=B(t,\bullet)+BU_1(\bullet,t)$ in the sector $\pi/4<\arg t<3\pi/4$. Using the relations $T_1=t^{3/2}U_1$ and $t=(\frac{4\tau}{i|\gamma|})^{1/4}$, one obtains
\begin{equation}\label{StokeConstantT1}
T_1(\tau)=U_1\left(\frac{4}{i|\gamma|}\right)^{2\rho}\Theta_1(\tau)(\bullet,\tau)+(\tau,\bullet),
\end{equation}
where we have chosen $B=(\frac{4}{i|\gamma|})^{\rho}$. Identifying Eq.\:\eqref{StokeConstantT1} with the definition of the Stokes constant $T_1$ in the complex $\tau$ plane, $T_1(\tau)=T_1\Theta_1(\tau)(\bullet,\tau)+(\tau,\bullet)$, one immediately obtains $T_1=U_1(\frac{4}{i|\gamma|})^{2\rho}$. Similarly, we have $T_2=U_2(\frac{4}{i|\gamma|})^{-2\rho}$, and in general $T_{2k+1}=U_{2k+1}(\frac{4}{i|\gamma|})^{2\rho}$ and $T_{2k}=U_{2k}(\frac{4}{i|\gamma|})^{-2\rho}$.

We now evaluate the transition probability $|U_1(\infty)|^2$ via the Stoke constants $U_k$. For later convenience, let us denote $A(\bullet,t)+B(t,\bullet)$ as the asymptotic WKB solution of $U_1(t)$ in the sector $7\pi/8<\arg t<\pi$. For $\gamma=|\gamma|$, one obtains
\begin{subequations}
\begin{align}
U_1(t)&=Ae^{i\theta(t)}+Bt^{-3}e^{-i\theta(t)},\label{WKBU1Gamma>0}\\
\dot{U}_1(t)&=\frac{iA}{2}(\gamma t^3+\alpha t)e^{i\theta(t)}-\frac{iB}{2}(\gamma+\alpha t^{-2})e^{-i\theta(t)},\label{WKBU1tGamma>0}
\end{align}
\end{subequations}
where $\theta(t)\equiv \frac{\gamma}{8}t^4+\frac{\alpha}{4}t^2$. Substituting the conditions $|U_1(-\infty)|=0$ and $|\dot{U}_1(-\infty)|=|f|$ into Eqs.\:\eqref{WKBU1Gamma>0} and \eqref{WKBU1tGamma>0}, one obtains $A=0$ and $B=2|f|/\gamma$. Similarity, for $\gamma=-|\gamma|$, one obtains
\begin{subequations}
\begin{align}
U_1(t)&=At^{-3}e^{-i\theta(t)}+Be^{i\theta(t)},\\
\dot{U}_1(t)&=\frac{iA}{2}(|\gamma|-\alpha t^{-2})e^{-i\theta(t)}-\frac{iB}{2}(|\gamma|t^3-\alpha t)e^{i\theta(t)},
\end{align}
\end{subequations}
which yield $B=0$ and $A=2|f|/|\gamma|$. For a clockwise crossing, the asymptotic WKB solution of $U_1(t)$ in the sector $0<\arg t<\pi/8$ has the form
\begin{align}
U_1(t)&=\{A[1+U_1U_2+U_4(U_1+U_3+U_1U_2U_3)]\nonumber\\
&-B(U_1+U_3+U_1U_2U_3)\}(\bullet,t)+[B(1+U_2U_3)\nonumber\\
&-A(U_2+U_4+U_2U_3U_4)](t,\bullet).
\end{align}
Hence, $|U_1(\infty)|=\frac{2|f|}{\gamma}|U_1+U_3+U_1U_2U_3|$ for $\gamma=|\gamma|$; whereas $|U_1(\infty)|=\frac{2|f|}{|\gamma|}|U_2+U_4+U_2U_3U_4|$ for $\gamma=-|\gamma|$. As a result, using the relations $T_{2k+1}=U_{2k+1}(\frac{4}{i|\gamma|})^{2\rho}$ and $T_{2k}=U_{2k}(\frac{4}{i|\gamma|})^{-2\rho}$, the final transition probability can be expressed via the Stokes constants $T_k$ as
\begin{subequations}
\begin{align}
|U_1(\infty)|^2&=\frac{|f|^2}{2\gamma^{1/2}}|T_1+T_3+T_1T_2T_3|^2,\:(\gamma=|\gamma|);\\
|U_1(\infty)|^2&=\frac{|f|^2}{2|\gamma|^{1/2}}|T_2+T_4+T_2T_3T_4|^2,\:(\gamma=-|\gamma|).
\end{align}
\end{subequations}
In appendix \ref{A}, we will show that the Stokes constants $T_k$ for the asymptotic WKB solutions of $T_1(\tau)$ satisfy a set of simple inter-relations, $T_{2k}=\bar{T}_{2k-1}e^{2i\pi\rho}$ and $T_{2k+1}=\bar{T}_{2k}e^{-2i\pi\rho}$, where $\bar{T}_k(Q_1,Q_2,Q_3)\equiv T_k(-iQ_1,-Q_2,iQ_3)$. In appendix \ref{B}, we will express the Stokes constant $T_1$ into a convergent series as a function of the parameters of the nonlinear detuning and the Rabi frequency. Hence, the final transition probability may also be expressed in an analytical form in terms of $\alpha$, $\gamma$ and $f$.

\section{Analytical Approximations to the transition amplitude}\label{IV}
As we have shown in the last section, the final transition probability $|U_1(t\rightarrow\infty)|^2$ can be expressed in an analytical form of a convergent series as a function of $\alpha$, $\gamma$ and $f$. However, a shortcoming of the explicit series solution is that the recursive relation for the coefficients of the series involve many terms that should be summed before obtaining a quantitative estimate. In this section, we provide a different approach based on the single assumption that the wave amplitude may be well-approximated by simple special functions in different physical limits. The main idea is that, after a suitable transformation $U_1(t)\rightarrow W_1(x)$, the differential equation which governs the wave amplitude becomes $W_1^{\prime\prime}+(F(x)+G(x))W_1=0$, where $G(x)$ may be neglected in that particular physical limit, and $W_1^{\prime\prime}+F(x)W_1=0$ may be solved by simple special functions. From Eq.\:\eqref{SchrodingerEquationWt}, for $x\equiv t^2\rightarrow \infty$, one obtains
\begin{equation}\label{XInfinity}
    F=\frac{\alpha^2}{16}-\frac{3i\gamma}{8}+\frac{\alpha\gamma}{8}x+\frac{\gamma^2}{16}x^2,\:
    G=\frac{3}{16}\frac{1}{x^2}+\left(\frac{|f|^2}{4}-\frac{i\alpha}{8}\right)\frac{1}{x};
\end{equation}
whereas for $x\rightarrow 0$, one obtains
\begin{equation}\label{XZero}
    F=\frac{3}{16}\frac{1}{x^2}+\left(\frac{|f|^2}{4}-\frac{i\alpha}{8}\right)\frac{1}{x}+\frac{\alpha^2}{16}-\frac{3i\gamma}{8},\:G=\frac{\alpha\gamma}{8}x+\frac{\gamma^2}{16}x^2.
\end{equation}
So far, the analysis is exact. But for $x\rightarrow \infty$, after neglecting $G(x)$, we obtain $W_1=c_1^{\RN{1}}U(a,\xi)+c_2^{\RN{1}}V(a,\xi)$, where $U(a,\xi)$ and $V(a,\xi)$ are the parabolic cylinder functions which satisfy $\frac{d^2W_1}{d\xi^2}-(\frac{1}{4}\xi^2+a)W_1=0$ with $a\equiv \frac{3}{4}$ and $\xi\equiv \sqrt{\frac{i\gamma}{2}}(x+\frac{\alpha}{\gamma})$, and $c_1^{\RN{1}}$ and $c_2^{\RN{1}}$ are two constants that will be determined later. Similarly, for $x\rightarrow 0$, after neglecting $G(x)$, we obtain $W_1=c_1^{\RN{2}}M_{\kappa,\mu}(\xi)+c_2^{\RN{2}}M_{\kappa,-\mu}(\xi)$, where $M_{\kappa,\pm\mu}(\xi)$ are the Whittaker functions which satisfy $\frac{d^2W_1}{d\xi^2}+(-\frac{1}{4}+\frac{\kappa}{\xi}+\frac{\frac{1}{4}-\mu^2}{\xi^2})W_1=0$ with $\xi\equiv \sqrt{\frac{3i\gamma}{2}-\frac{\alpha^2}{4}} x$, $\kappa\equiv (\frac{3i\gamma}{2}-\frac{\alpha^2}{4})^{-\frac{1}{2}}(\frac{|f|^2}{4}-\frac{i\alpha}{8})$ and $\mu\equiv 1/4$, and $c_1^{\RN{2}}$ and $c_2^{\RN{2}}$ are two constants that will be determined later. In order to provide analytical approximations to $W_1(x)$, we assume that the solution $c_1^{\RN{1}}U(a,\xi)+c_2^{\RN{1}}V(a,\xi)$ is approximately valid for $x\geq x^*$ (region I); whereas the solution $c_1^{\RN{2}}M_{\kappa,\mu}(\xi)+c_2^{\RN{2}}M_{\kappa,-\mu}(\xi)$ is approximately valid for $x\leq x^*$ (region II). Here, $x^*$ is a transition point that will be determined later. 
\begin{figure*}
	\subfloat[$|f|=\alpha=1$,$\gamma=4$\label{sfig:SuperHeunApprox1Beta4}]{%
	\includegraphics[width=0.8\columnwidth]{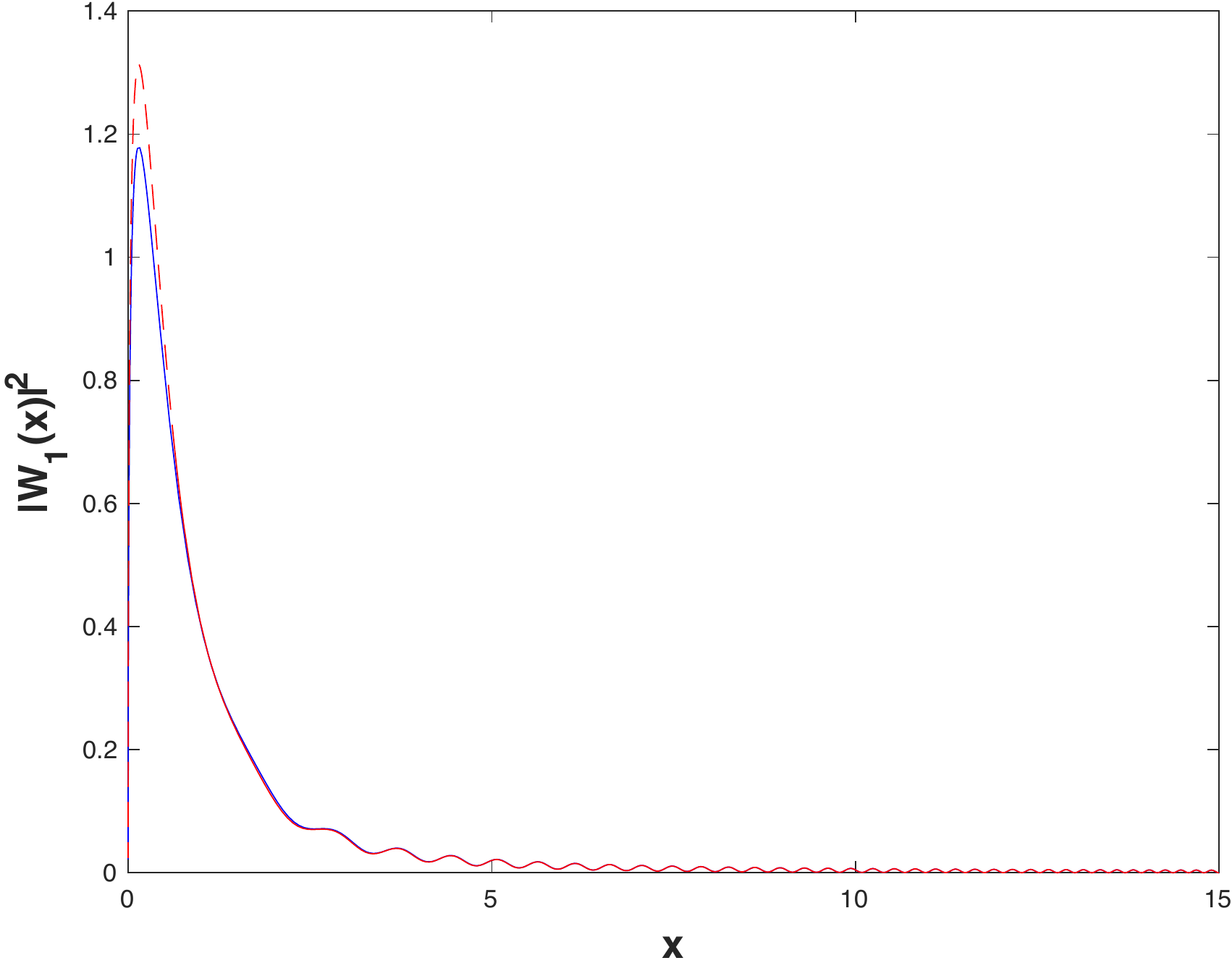}%
	}
	\subfloat[$|f|=\alpha=1$,$\gamma=2$\label{sfig:SuperHeunApprox1Beta2}]{%
 	 \includegraphics[width=0.8\columnwidth]{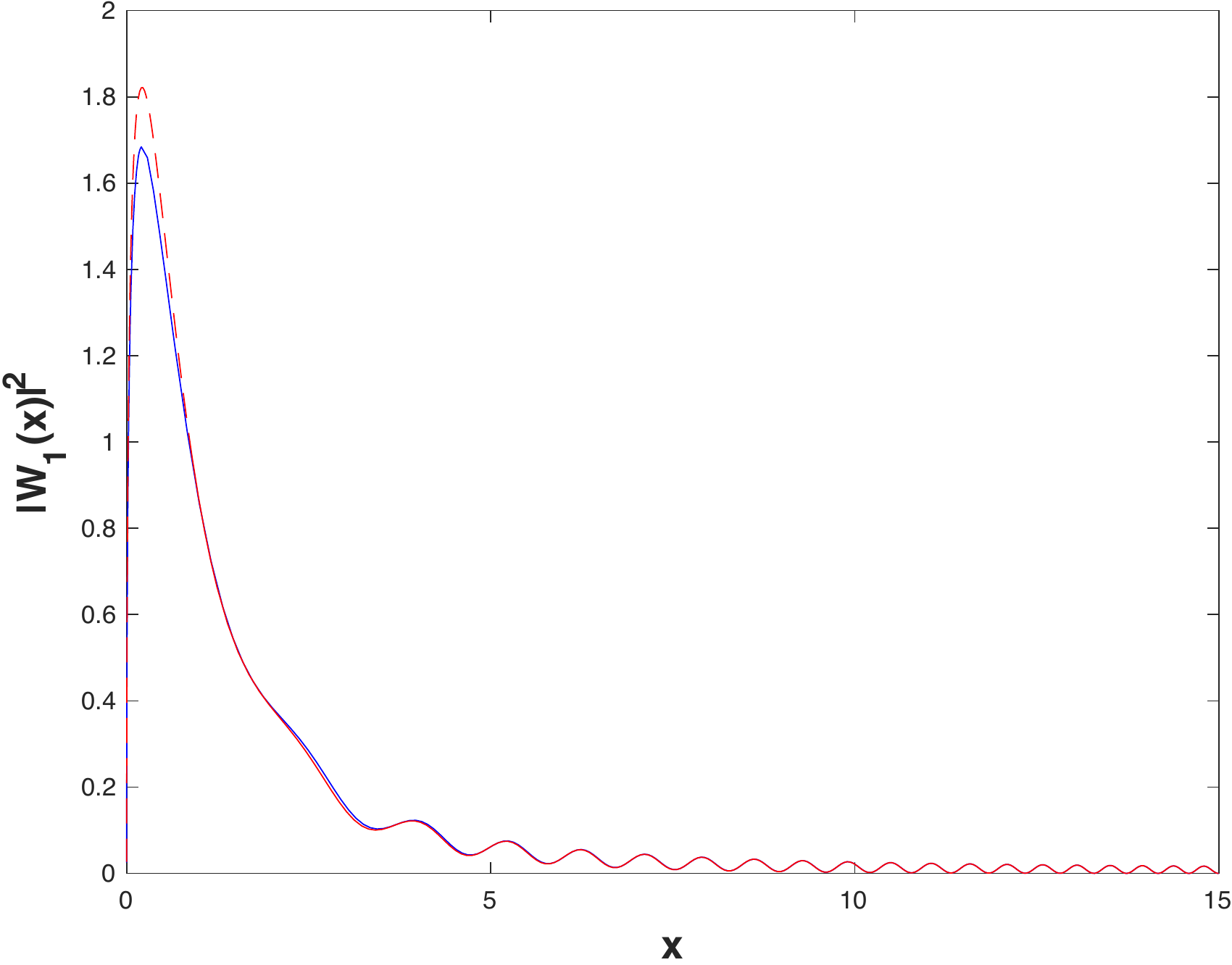}%
	}\hfill
	\subfloat[$|f|=\alpha=1$,$\gamma=1$\label{sfig:BiConfluentHeunApprox(Gamma1)}]{%
	\includegraphics[width=0.8\columnwidth]{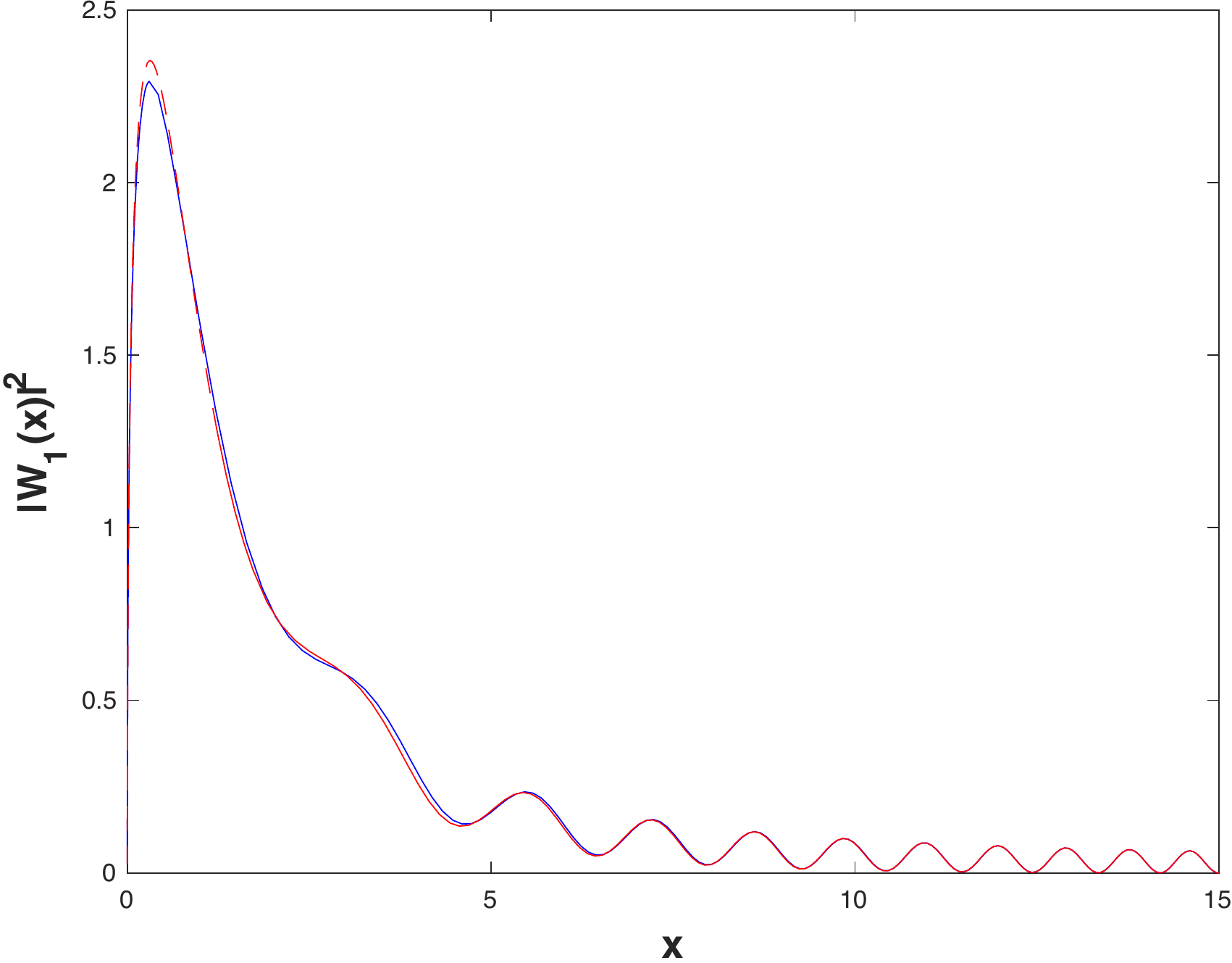}%
	}
	\subfloat[$|f|=\alpha=1$,$\gamma=0.5$\label{sfig:BiConfluentHeunApprox(Gamma05)}]{%
 	 \includegraphics[width=0.8\columnwidth]{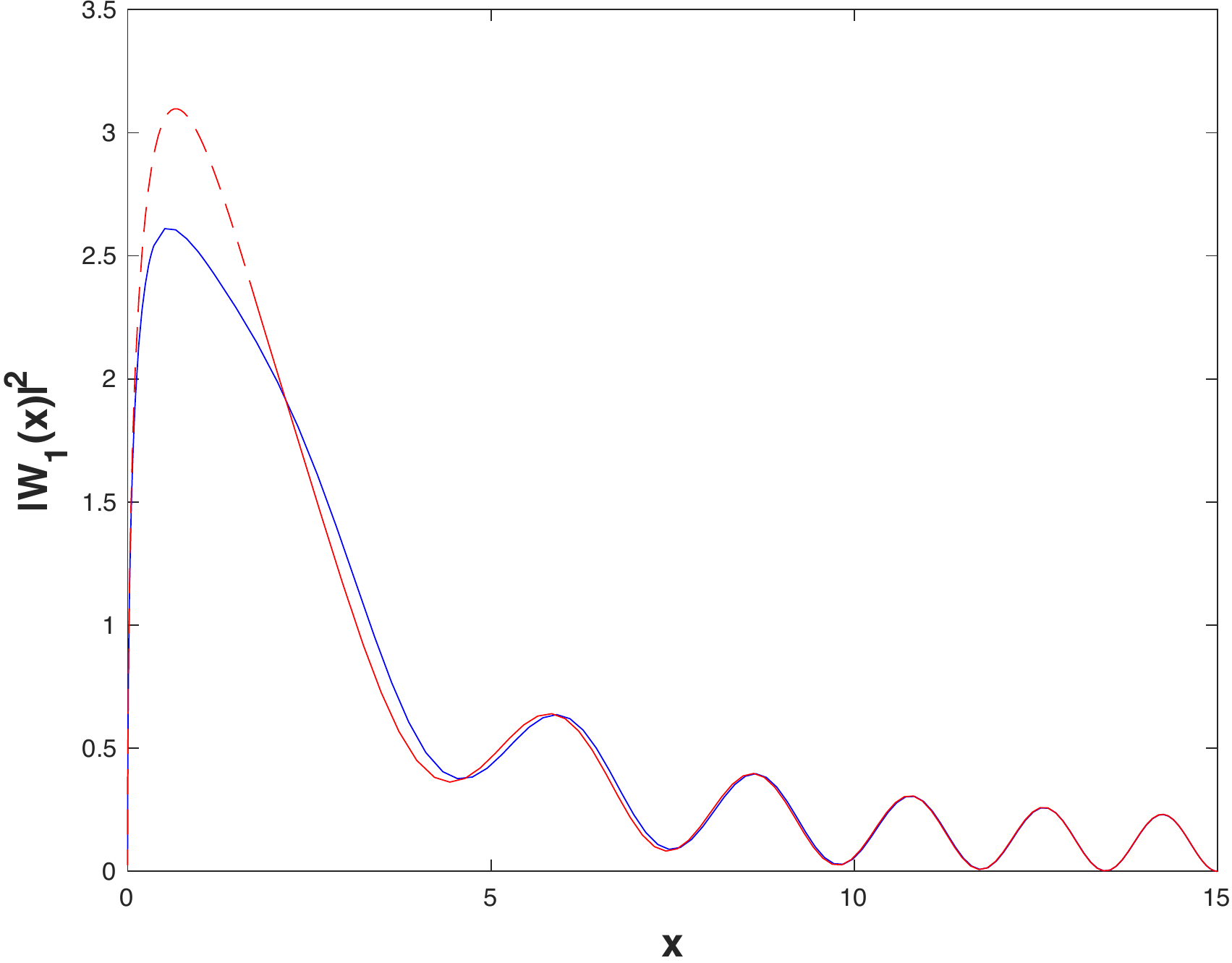}%
	}
\caption{Comparison of approximate and numerical solutions of $|W_1(x)|^2$. Numerical solutions are plotted in blue solid lines, and the approximate solutions are plotted in red lines.}
\label{fig:SuperHeunApprox}
\end{figure*}

\begin{figure*}
	\subfloat[$|f|=\alpha=1$,$\gamma=1$\label{sfig:SuperparabolicC01}]{%
	\includegraphics[width=\columnwidth]{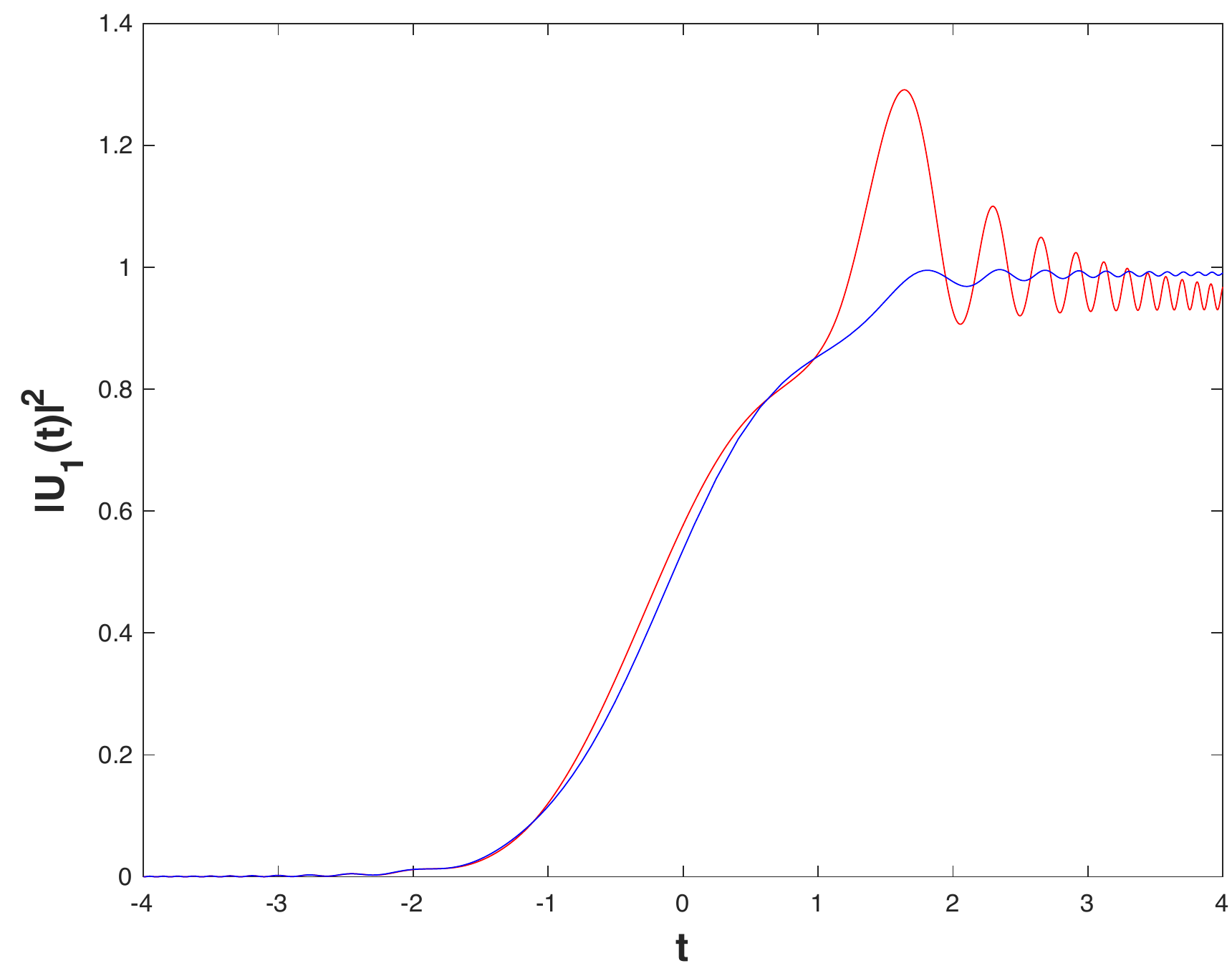}%
	}\hfill
	\subfloat[$|f|=\alpha=1$,$\gamma=1.5$\label{sfig:SuperparabolicC015}]{%
 	 \includegraphics[width=\columnwidth]{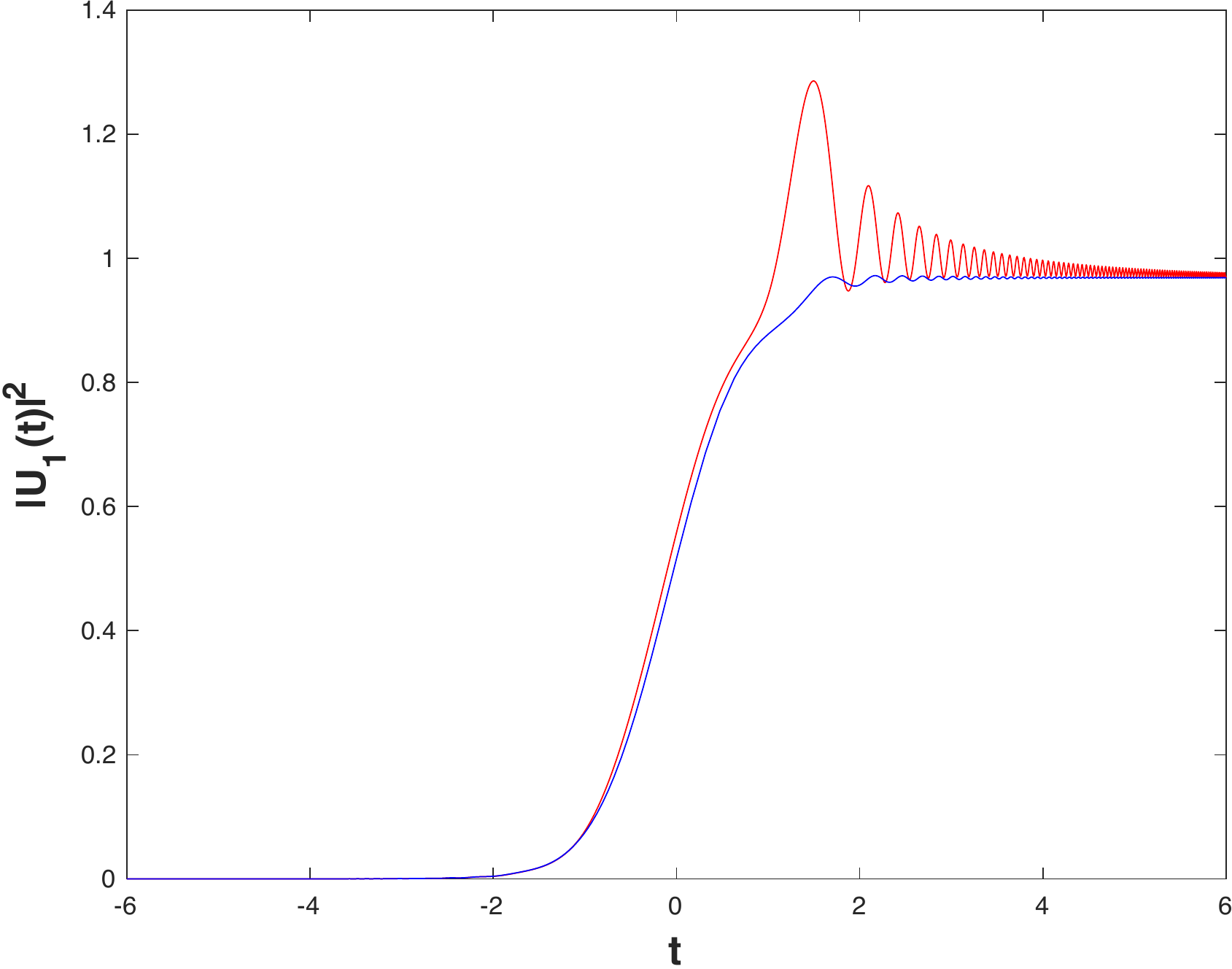}%
	}
\caption{Comparison of approximate and numerical solutions of the transition probability $|U_1(t)|^2$. Numerical solutions are plotted in blue solid lines, and the approximate solutions are plotted in red lines.}
\label{fig:SuperParabolicEnd}
\end{figure*}

One may determine the transition point $x^*$ via evaluation of errors from neglecting $G(x)$ in Eqs.\:\eqref{XInfinity} and \eqref{XZero} (see appendix \ref{C}). But here we determine the transition point $x^*$ in a simpler way --- we minimize the maximal error between the approximate and real solutions. For $x\geq x^*$, the bi-confluent Heun function $W_1(x)$ can be approximated by $c_1^{\RN{1}}U_a(\xi^{\RN{1}})+c_2^{\RN{1}}V_a(\xi^{\RN{1}})$ with $a\equiv\frac{3}{4}$ and $\xi^{\RN{1}}\equiv \sqrt{\frac{i\gamma}{2}}(x+\frac{\alpha}{\gamma})$, where $c_1^{\RN{1}}$ and $c_2^{\RN{1}}$ are two constants determined by the initial conditions
\begin{subequations}
\begin{align}
W_1(x_0)&=c_1^{\RN{1}}U(a,\xi^{\RN{1}}(x_0))+c_2^{\RN{1}}V(a,\xi^{\RN{1}}(x_0)),\\
W_1^\prime(x_0)&=c_1^{\RN{1}}U^\prime(a,\xi^{\RN{1}}(x_0))+c_2^{\RN{1}}V^\prime(a,\xi^{\RN{1}}(x_0)).
\end{align}
\end{subequations}
For $x\leq x^*$, the bi-confluent Heun function $W_1(x)$ can be approximated by $c_1^{\RN{2}}M_{\kappa,\mu}(\xi^{\RN{2}})+c_2^{\RN{2}}M_{\kappa,-\mu}(\xi^{\RN{2}})$ with $\xi^{\RN{2}}\equiv \sqrt{-a_2}x$, $\kappa\equiv\frac{a_0}{4\sqrt{-a_2}}$ and $\mu\equiv\frac{1}{4}$, where $a_2\equiv \frac{\alpha^2}{4}-\frac{3i\gamma}{2}$ and $a_0\equiv |f|^2-\frac{i\alpha}{2}$. Here $c_1^{\RN{2}}$ and $c_2^{\RN{2}}$ are two constants determined by requiring both $W_1(x)$ and $W_1^\prime(x)$ are continuous at $x^*$
\begin{subequations}
\begin{align}
   c_1^{\RN{1}}U_a(x^*)+c_2^{\RN{1}}V_a(x^*)&=c_1^{\RN{2}}M_{\kappa,\mu}(x^*)+c_2^{\RN{2}}M_{\kappa,-\mu}(x^*),\\
   c_1^{\RN{1}}U_a^\prime(x^*)+c_2^{\RN{1}}V_a^\prime(x^*)&=c_1^{\RN{2}}M^\prime_{\kappa,\mu}(x^*)+c_2^{\RN{2}}M^\prime_{\kappa,-\mu}(x^*),
\end{align}
\end{subequations}
where $U_a(x^*)$, $V_a(x^*)$, $M_{\kappa,\mu}(x^*)$ and $M_{\kappa,-\mu}(x^*)$ are shorthands for $U_a(\xi^{\RN{1}}(x^*))$, $V_a(\xi^{\RN{1}}(x^*))$, $M_{\kappa,\mu}(\xi^{\RN{2}}(x^*))$ and $M_{-\kappa,\mu}(\xi^{\RN{2}}(x^*))$ respectively. Now, the two constants $c_1^{\RN{2}}(x^*)$ and $c_2^{\RN{2}}(x^*)$ are function of $x^*$. Hence, the maximal error between the approximate and real solutions as a function of $x^*$ is
\begin{equation}
    \mathcal{E}(x^*)\equiv\max_{x\in(0,x^*)}|W_1(x)-[c_1^{\RN{2}}(x^*)M^\prime_{\kappa,\mu}(x)+c_2^{\RN{2}}(x^*)M^\prime_{\kappa,-\mu}(x)]|.
\end{equation}
The actual transition point $\bar{x}^*$ is then chosen by minimizing the maximal error function $\mathcal{E}(x^*)$
\begin{equation}
    \mathcal{E}(\bar{x}^*)=\min_{x^*\in(0,\infty)}\mathcal{E}(x^*).
\end{equation}
In Fig.\:\ref{fig:SuperHeunApprox}, we compare the analytical approximations to $W_1(x)$ with numerical results of simulations. The result shows that $W_1(x)$ may be well-approximated by parabolic cylinder functions and Whittaker functions in both long and short time limits, as long as the maximal error function $\mathcal{E}(x^*)$ is minimized. In Fig.\:\ref{fig:SuperParabolicEnd}, after transforming back to $U_1(t)$, we plot the analytical approximations to the transition probability $|U_1(t)|^2$, and compare it with numerical simulations. The result shows that the transition probability $|U_1(t)|^2$ may be well-approximated by the above-mentioned special functions in a large part of parameter space. 

\section{Conclusion}\label{V}
In this work, we discussed the transition dynamics of a two-level crossing model in which the linear Landau-Zener detuning is modified by a cubic nonlinearity near the crossing point. We studied the final transition probability by use of two different approaches. One approach is to use the asymptotic WKB solutions to derive a closed-form series expression of the final transition probability. Another approach is to connect local analytical approximations to the transition amplitudes in different physical limits. The analytical approximations is validated via comparison with numerical results of simulations. In this work, we only discuss the cases that the parameters of the detuning are all real. In future works, we may extend our current work to non-Hermitian systems \cite{longstaff2019non} in which the parameters of the detuning are in general complex.

\begin{acknowledgements}
The Authors would like to thank the Science and Technology Development Fund of the Macau SAR for providing support, FDCT 023/2017/A1.
\end{acknowledgements}

\begin{appendix}
\section{Inter-relations between the Stokes constants $T_k$}\label{A}
In this appendix, we derive the inter-relations between the various Stokes constants $T_k$ in the complex $\tau$ plane. To begin with, let us recall the Schr\"{o}dinger equation which governs the wave amplitude $T_1(\tau)$: $T_1^{\prime\prime}(\tau)=\left(\frac{1}{4}+\sum_{n=1}^4Q_n\tau^{-\frac{n}{2}}\right)T_1(\tau)$, where $Q_n$ are some coefficients given by Eq.\:\eqref{Qn}. For $\pi/2<\arg\tau<\pi$, one may denote the general solution for the wave amplitude as $T_1(\tau)=Au(\tau;Q_n) + Bv(\tau;Q_n)$, where $u(\tau;Q_n)$ and $v(\tau;Q_n)$ are two independent solutions given by Eq.\:\eqref{IndependentUandV}. In particular, we consider the solution with $A=0$ and $B=1$, then we obtain $T_1^{(1)}(\tau)=v(\tau;Q_n)$ and
\begin{subequations}
\begin{align}\label{StokesOnComplexTauPlane1}
    T_1^{(2)}&=T_1\Theta_1 u+v,T^{(3)}=T_1\Theta_1 u+(1+T_1T_2\Theta_2)v,\\
    T_1^{(4)}&=(T_1\Theta_1+(1+T_1T_2)T_3\Theta_3)u+(1+T_1T_2\Theta_2)v,\\
    T_1^{(5)}&=T_1^{(4)}+(T_1+T_3+T_1T_2T_3)T_4\Theta_4v,\\
    T_1^{(6)}&=T_1^{(5)}+(1+T_1T_2\nonumber\\
    &+(T_1+T_3+T_1T_2T_3)T_4)T_5\Theta_5u,\\
    T_1^{(7)}&=T_1^{(6)}+[T_1+(1+T_1T_2)T_3\nonumber\\
    &+(1+T_1T_2+(T_1+T_3+T_1T_2T_3)T_4)T_5]T_6\Theta_6 v,\\
    T_1^{(8)}&=T_1^{(7)}+\{1+T_1T_2+(T_1+T_3+T_1T_2T_3)T_4\nonumber\\
    &+[T_1+(1+T_1T_2)T_3+(1+T_1T_2\nonumber\\
    &+(T_1+T_3+T_1T_2T_3)T_4)T_5]T_6\}T_7\Theta_7u,\\
    T_1^{(9)}&=T_1^{(8)}+\{T_1+(1+T_1T_2)T_3\nonumber\\
    &+(1+T_1T_2+(T_1+T_3+T_1T_2T_3)T_4)T_5)\nonumber\\
    &+[1+T_1T_2+(T_1+T_3+T_1T_2T_3)T_4\nonumber\\
    &+(T_1+(1+T_1T_2)T_3+(1+T_1T_2\nonumber\\
    &+(T_1+T_3+T_1T_2T_3)T_4)T_5)T_6]T_7\}T_8\Theta_8v,\label{StokesOnComplexTauPlane7}
\end{align}
\end{subequations}
where $\Theta_k(\tau)$ is the step function, which equals to $0$ for $\arg\tau\leq k\pi$, and equals to $1$ for $\arg\tau>k\pi$, and $T_1^{(k)}(\tau)=A^{(k)}u+B^{(k)}v$ denotes the solution for $\pi/2<\arg\tau<k\pi$. From Eqs.\:\eqref{StokesOnComplexTauPlane1} - \eqref{StokesOnComplexTauPlane7}, we see that the relations between $T_1^{(k)}(\tau)$ and $T_1^{(k-1)}(\tau)$ are
\begin{subequations}
\begin{align}
    T_1^{(k)}&=T_1^{(k-1)}+A^{(k-1)}T_{k-1}\Theta_{k-1}v,\:(\mbox{for}\:k\:\mbox{odd});\\
    T_1^{(k)}&=T_1^{(k-1)}+B^{(k-1)}T_{k-1}\Theta_{k-1}u,\:(\mbox{for}\:k\:\mbox{even}).
\end{align}
\end{subequations}
As $\arg(e^{-i\pi}\tau)=\arg\tau-\pi$, we have $\Theta_k(e^{-i\pi}\tau)=\Theta_{k+1}(\tau)$. Using the relations $\bar{u}=e^{i\pi\rho}v$ and $\bar{v}=e^{-i\pi\rho}u$, we obtain
\begin{subequations}
\begin{align}\label{BarA}
    \bar{T}_1^{(1)}&=\bar{T}_1^{(2)}=e^{-i\pi\rho}u,
    \bar{T}^{(3)}=e^{-i\pi\rho}u+\bar{T}_1\Theta_2 e^{i\pi\rho}v,\\
    \bar{T}_1^{(4)}&=(1+\bar{T}_1\bar{T}_2\Theta_3)e^{-i\pi\rho}u+\bar{T}_1\Theta_2e^{i\pi\rho}v,\\
    \bar{T}_1^{(5)}&=\bar{T}_1^{(4)}+(1+\bar{T}_1\bar{T}_2)\bar{T}_3\Theta_4e^{i\pi\rho}v,\\
    \bar{T}_1^{(6)}&=\bar{T}_1^{(5)}+(\bar{T}_1+\bar{T}_3+\bar{T}_1\bar{T}_2\bar{T}_3)\bar{T}_4\Theta_5e^{-i\pi\rho}u,\\
    \bar{T}_1^{(7)}&=\bar{T}_1^{(6)}+(1+\bar{T}_1\bar{T}_2\nonumber\\
    &+(\bar{T}_1+\bar{T}_3+\bar{T}_1\bar{T}_2\bar{T}_3)\bar{T}_4)\bar{T}_5\Theta_6)e^{i\pi\rho}v,\\
    \bar{T}_1^{(8)}&=\bar{T}_1^{(7)}+[\bar{T}_1+(1+\bar{T}_1\bar{T}_2)\bar{T}_3+(1+\bar{T}_1\bar{T}_2\nonumber\\
    &+(\bar{T}_1+\bar{T}_3+\bar{T}_1\bar{T}_2\bar{T}_3)\bar{T}_4)\bar{T}_5]\bar{T}_6\Theta_7 e^{-i\pi\rho}u,\\
    \bar{T}_1^{(9)}&=\bar{T}_1^{(8)}+\{1+\bar{T}_1\bar{T}_2+(\bar{T}_1+\bar{T}_3+\bar{T}_1\bar{T}_2\bar{T}_3)\bar{T}_4\nonumber\\
    &+[\bar{T}_1+(1+\bar{T}_1\bar{T}_2)\bar{T}_3+(1+\bar{T}_1\bar{T}_2\nonumber\\
    &+(\bar{T}_1+\bar{T}_3+\bar{T}_1\bar{T}_2\bar{T}_3)\bar{T}_4)\bar{T}_5]\bar{T}_6\}\bar{T}_7\Theta_8e^{i\pi\rho}v,\label{BarG}
\end{align}
\end{subequations}
where $\bar{T}_k(Q_1,Q_2,Q_3)\equiv T_k(-iQ_1,-Q_2,iQ_3)$. We now consider the solution for $\pi/2<\arg\tau<\pi$ with $A=e^{-i\pi\rho}$ and $B=0$, where $\rho\equiv Q_2-Q_1^2=\pm\frac{3}{8}$ for $\gamma=\pm|\gamma|$. Then we obtain $T_1^{(1)}=T_1^{(2)}=e^{-i\pi\rho}u$ and
\begin{subequations}
\begin{align}\label{NoBarA}
    T_1^{(3)}&=T_1^{(2)}+e^{-i\pi\rho}T_2\Theta_2v,\\
    T_1^{(4)}&=T_1^{(3)}+e^{-i\pi\rho}T_2T_3\Theta_3u,\\
    T_1^{(5)}&=T_1^{(4)}+e^{-i\pi\rho}(1+T_2T_3)T_4\Theta_4v,\\
    T_1^{(6)}&=T_1^{(5)}+e^{-i\pi\rho}(T_2+(1+T_2T_3)T_4)T_5\Theta_5u,\\
    T_1^{(7)}&=T_1^{(6)}+e^{-i\pi\rho}[1+T_2T_3\nonumber\\
    &+(T_2+(1+T_2T_3)T_4)T_5]T_6\Theta_6v,\\
    T_1^{(8)}&=T_1^{(7)}+e^{-i\pi\rho}[T_2+(1+T_2T_3)T_4+(1+T_2T_3\nonumber\\
    &+(T_2+(1+T_2T_3)T_4)T_5)T_6]T_7\Theta_7 u,\\
    T_1^{(9)}&=T_1^{(8)}+e^{-i\pi\rho}\{1+T_2T_3+(T_2+(1+T_2T_3)T_4)T_5\nonumber\\
    &+[T_2+(1+T_2T_3)T_4+(1+T_2T_3\nonumber\\
    &+(T_2+(1+T_2T_3)T_4)T_5)T_6]T_7\}T_8\Theta_8 v,\label{NoBarG}
\end{align}
\end{subequations}
By comparision, Eqs.\:\eqref{BarA} - \eqref{BarG} and \eqref{NoBarA} - \eqref{NoBarG} imply the following inter-relations between the Stokes constants $T_k$
\begin{align}
    T_2&=\bar{T}_1e^{2i\pi\rho},
    T_3=\bar{T}_2e^{-2\pi i\rho},
    T_4=\bar{T}_3e^{2i\pi\rho},
    T_5=\bar{T}_4e^{-2i\pi\rho},\nonumber\\
    T_6&=\bar{T}_5e^{2i\pi\rho},
    T_7=\bar{T}_6e^{-2i\pi\rho},
    T_8=\bar{T}_7e^{2i\pi\rho}.
\end{align}

\section{Explicit expression for the Stoke constant $T_1$}\label{B}
In this appendix, we use Hinton's integral equation method \cite{hinton1979stokes, zhu1992stokes} to derive an explicit formula for the Stoke constant $T_1$ in the complex $\tau$ plane. To begin with, let us express the solution of ${T}_1^{\prime\prime}(\tau)+Q(\tau)T_1(\tau)=0$ as $T_1(\tau)=\mu(\tau)^{-1/2}[A_1(\tau)e^{\int\mu(\tau)d\tau}+A_2(\tau)e^{-\int\mu(\tau)d\tau}]$, we then obtain
\begin{align}
T_1^{\prime\prime}&=\mu^{-1/2}\{[A^{\prime\prime}_1+2(\mu-\chi)A^\prime_1+(\chi^2-\chi^\prime+\mu^2)A_1]e^{\int\mu d\tau}\nonumber\\
&+[A^{\prime\prime}_2-2(\mu+\chi)A^\prime_2+(\chi^2-\chi^\prime+\mu^2)A_2]e^{-\int\mu d\tau}\},\label{SecondOrderDerivative}
\end{align}
where $\chi\equiv\mu^\prime/(2\mu)$. One may substitute the ansatzes $A^\prime_1=Be^{-2\int \mu d\tau}A_2$ and $A^\prime_2=Be^{2\int \mu d\tau}A_1$ into Eq.\:\eqref{SecondOrderDerivative} and $T^{\prime\prime}_1(\tau)+Q(\tau)T_1(\tau)=0$, and obtain a Riccati equation
\begin{equation}
    v^\prime+v^2+\mu^2+Q=0,
\end{equation}
where $v=B-\chi$. Let us denote $v=X^\prime/X$, we then obtain $X^{\prime\prime}+(\mu^2+Q)X=0$. The function $\mu(\tau)$ is fixed by requiring $X(\tau)$ has a regular singularity at $\tau=\infty$. In our case, $Q(\tau)=-(\frac{1}{4}+\sum_{n=1}^4Q_n\tau^{-\frac{n}{2}})$, which yields $\mu(\tau)=\frac{1}{2}+\sum_{n=1}^3\mu_n\tau^{-\frac{n}{2}}$ with $\mu_1\equiv Q_1$, $\mu_2\equiv Q_2-Q_1^2$ and $\mu_3\equiv Q_3-2Q_1(Q_2-Q_1^2)$. Hence, we obtain
\begin{equation}
    \int\mu(\tau)d\tau = \frac{\tau}{2}+2\mu_1\tau^{1/2}+\mu_2\ln\tau-2\mu_3\tau^{-1/2},
\end{equation}
and $\mu^2(\tau)+Q(\tau)=\sum_{n=1}^3P_n\tau^{-\frac{n+3}{2}}$, where
\begin{align}
    P_1&\equiv(Q_2-Q_1^2)^2+2Q_1Q_3-4Q_1^2(Q_2-Q_1^2)-Q_4,\nonumber\\
    P_2&\equiv2(Q_2-Q_1^2)[Q_3-2Q_1(Q_2-Q_1^2)],\nonumber\\
    P_3&\equiv[Q_3-2Q_1(Q_2-Q_1^2)]^2.
\end{align}
Thus, $(\bullet,\tau)\equiv\mu^{-1/2}e^{\int\mu d\tau}$ and $(\tau,\bullet)\equiv\mu^{-1/2}e^{-\int\mu d\tau}$ are dominant and sub-dominant WKB solutions for $T_1(\tau)$ on $\Re{\tau}>0$ respectively. We now expand $\chi(\tau)$ in a series as $\sum_{n=1}^\infty\chi_n\tau^{-\frac{n+1}{2}}$ with $\chi_1=-\frac{1}{4}\mu_1$, $\chi_2=\frac{1}{4}\mu_1^2-\frac{1}{2}\mu_2$, $\chi_3=-\frac{1}{4}\mu_1^3+\frac{3}{4}\mu_1\mu_2-\frac{3}{4}\mu_3$, and $\chi_n=-\sum_{m=1}^3\mu_m\chi_{n-m}$ for $n>3$. Similarly, we may expand $v(\tau)$ in a series as $\sum_{n=1}^\infty v_n\tau^{-\frac{n+1}{2}}$, where the coefficients $v_n$ obey $v_1^2-v_1+P_1=0$, $v_2=P_2/(3/2-2v_1)$ and $v_n=(\sum_{m=2}^{n-1}v_mv_{n+1-m}+P_n)/(\frac{n+1}{2}-2v_1)$ for $n>2$. Now, in order to derive the explicit expression of the Stoke constant $T_1$, we consider a sub-dominant solution for $T_1(\tau)$ on $0<\arg\tau<\pi/2$. Assuming $A_1(\tau)\rightarrow 0$ and $A_2(\tau)\rightarrow 1$ on the Stoke line $\arg\tau=0$, one obtains
\begin{subequations}
\begin{align}
    A_1(\tau)&=\int_\infty^\tau B_1(s)e^{-s-4\mu_1s^{1/2}}s^{-2\mu_2}A_2(s)ds,\label{A1definition}\\
    A_2(\tau)&=1+\int_\infty^\tau B_2(s)e^{s+4\mu_1s^{1/2}}s^{2\mu_2}A_1(s)ds\label{A2definition},
\end{align}
\end{subequations}
where $B_1(s)\equiv B(s)e^{4\mu_3s^{-\frac{1}{2}}}=\sum_{n=1}^\infty B^{(1)}_ns^{-\frac{n+1}{2}}$, and $B_2(s)=B(s)e^{-4\mu_3s^{-\frac{1}{2}}}=\sum_{n=1}^\infty B^{(2)}_ns^{-\frac{n+1}{2}}$. The coefficients $B_n^{(1)}$ and $B_n^{(2)}$ obey
\begin{equation}
    B_n^{(1)}=\sum_{m=0}^{n-1}B_{n-m}\frac{(4\mu_3)^m}{m!},\:
    B_n^{(2)}=\sum_{m=0}^{n-1}B_{n-m}\frac{(-4\mu_3)^m}{m!},
\end{equation}
where $B_n\equiv \chi_n+v_n$. For $\tau\rightarrow\infty$, we have $T_1(\tau)=(\tau,\bullet)_s$ on $0<\arg\tau<\pi/2$, $T_1(\tau)=(\tau,\bullet)_d$ on $\pi/2<\arg\tau<\pi$, $T_1(\tau)=(\tau,\bullet)_d+T_1(0,\tau)_s$ on $\pi<\arg\tau<3\pi/2$, and $T_1(\tau)=(\tau,\bullet)_s+T_1(\bullet,\tau)_d$ on $3\pi/2<\arg\tau<2\pi$. Hence, on $0<\arg\tau<2\pi$, we may write $A_1(\tau)=T_1\Theta_1(\tau)$ and $A_2(\tau)=1$ for $\tau\rightarrow\infty$, and obtain the following asymptotic solutions of $T_1(\tau)$
\begin{subequations}
\begin{align}
    A_1(\tau)&=\sum_{n=0}^\infty\left[ e^{-\tau-4\mu_1\tau^{1/2}}\tau^{-2\mu_2}\alpha_n^{(1)}+T_1\Theta_1(\tau)\beta_n^{(2)}\right]\tau^{-n/2},\label{A1expand}\\
    A_2(\tau)&=\sum_{n=0}^\infty\left[\beta_n^{(1)}+T_1\Theta_1(\tau)e^{\tau+4\mu_1\tau^{1/2}}\tau^{2\mu_2}\alpha_n^{(2)}\right]\tau^{-n/2},\label{A2expand}
\end{align}
\end{subequations}
where $\alpha_0^{(1)}=\alpha_0^{(2)}=0$ and $\beta_0^{(1)}=\beta_0^{(2)}=1$. In order to determine the coefficients $\alpha_n^{(1)}$, $\alpha_n^{(2)}$, $\beta_n^{(1)}$, $\beta_n^{(2)}$ and $T_1$, we may use the asymptotic expansion for the incomplete gamma function
\begin{align*}
    &\Gamma(\beta,\tau)\equiv\int_\tau^\infty e^{-\zeta}\zeta^{\beta-1}d\zeta\nonumber\\
    &=\sum_{n=0}^\infty[(-1)^ne^{-\tau}\frac{\Gamma(n+1-\beta)}{\Gamma(1-\beta)}\tau^{-n+\beta-1}+\frac{2\pi ie^{i\pi(\beta-1)}}{\Gamma(1-\beta)}\Theta_1(\tau)].
\end{align*}
Substitution of Eq.\:\eqref{A2expand} into Eq.\:\eqref{A1definition} yields
\begin{align}\label{A1Series}
    A_1(\tau)=&-\sum_{r=0}^\infty\Delta_r^{(1)}e^{-\tau}\sum_{n=0}^\infty\frac{d^n}{d\tau^n}(\tau^{-2\mu_2-1-\frac{r}{2}}e^{-4\mu_1\tau^{\frac{1}{2}}})\nonumber\\
    &+\sum_{r=0}^\infty\Delta_r^{(1)}\sum_{n=0}^\infty\frac{(-4\mu_1)^n}{n!}\frac{2\pi ie^{i\pi(\frac{n-r}{2}-2\mu_2)}}{\Gamma(1+\frac{r-n}{2}+2\mu_2)}\Theta_1(\tau)\nonumber\\
    &+T_1\Theta_1(\tau)\sum_{n=1}^\infty\beta_n^{(2)}\tau^{-\frac{n}{2}},
\end{align}
where $\Delta_n^{(1)}\equiv\sum_{p+q=n}B_{p+1}^{(1)}\beta_q^{(1)}$, $\beta_n^{(2)}\equiv -\frac{2}{n}\sum_{p+q=n-1}B_{p+1}^{(1)}\alpha_{q+1}^{(2)}$. Comparing Eq.\:\eqref{A1expand} and \eqref{A1Series}, we immediately obtain the series expansion of the Stoke constant $T_1$ in terms of $B_n^{(1)}$ and $\beta_n^{(1)}$
\begin{equation}
    T_1=\sum_{r=0}^\infty\Delta_r^{(1)}\sum_{n=0}^\infty\frac{(-4\mu_1)^n}{n!}\frac{2\pi ie^{i\pi(\frac{n-r}{2}-2\mu_2)}}{\Gamma(1+\frac{r-n}{2}+2\mu_2)}.
\end{equation}
Hence, for $0<\arg\tau<\pi$, we have
\begin{align}
    &A_1(\tau)=-\sum_{r=0}^\infty\Delta_r^{(1)}e^{-\tau-4\mu_1\tau^{\frac{1}{2}}}\tau^{-2\mu_2}\cdot\nonumber\\
    &\sum_{n=0}^\infty\sum_{m=0}^n\binom{n}{m}\frac{\Gamma(-2\mu_2-\frac{r}{2})\tau^{-n+m-1-\frac{r}{2}}}{\Gamma(-2\mu_2-\frac{r}{2}-n+m)}\cdot\nonumber\\
    &\sum_{p=1}^m\sum_{q=0}^p(-1)^q\frac{(4\mu_1)^p}{q!(p-q)!}\frac{\Gamma(\frac{q}{2}+1)}{\Gamma(\frac{q}{2}+1-m)}\tau^{\frac{p}{2}-m}.
\end{align}
If we exchange the summation with respect to $m$ and $p$, and perform the change of variable $s\equiv 2n-p$, and equivalently $p=2n-s$, we obtain
\begin{align}
    &\sum_{k=0}^\infty\alpha_k^{(1)}\tau^{-\frac{k}{2}}=-\sum_{r=0}^\infty\Delta_r^{(1)}\sum_{n=0}^\infty\sum_{s=n}^{2n}\sum_{m=2n-s}^n\binom{n}{m}\cdot\nonumber\\
    &\frac{\Gamma(-2\mu_2-\frac{r}{2})}{\Gamma(-2\mu_2-\frac{r}{2}-n+m)}c_{(2n-s)m}(4\mu_1)^{2n-s}\tau^{-\frac{r+s+2}{2}},
\end{align}
where
\begin{equation}
    c_{pm}\equiv\sum_{q=0}^p\frac{(-1)^q}{q!(p-q)!}\frac{\Gamma(\frac{q}{2}+1)}{\Gamma(\frac{q}{2}+1-m)},\:\:(\mbox{for}\:\:p\geq 1).
\end{equation}
Finally, if we exchange the summation with respect to $n$ and $s$, we obtain the recursion relations between $\alpha_n^{(1)}$ and $\beta_n^{(1)}$
\begin{equation}
\alpha_{k+2}^{(1)}=-\sum_{r+s=k}\Delta_r^{(1)}\sum_{n=\frac{s}{2}}^s\sum_{m=2n-s}^n\binom{n}{m}\frac{\Gamma(-2\mu_2-\frac{r}{2})c_{(2n-s)m}(4\mu_1)^{2n-s}}{\Gamma(-2\mu_2-\frac{r}{2}-n+m)},
\end{equation}
where $\alpha_1^{(1)}=\alpha_2^{(1)}=0$. To find the recursion relation between $\alpha_n^{(2)}$ and $\beta_n^{(2)}$, we use the relations $\int_{\infty}^\tau s^{\beta-1}e^s ds=\int_{-\infty}^\tau s^{\beta-1}e^s ds=e^{i\pi(\beta-1)}\Gamma(\beta,e^{-i\pi}\tau)$, which yields the following asymptotic expansion
\begin{equation}
\int_{\infty}^\tau s^{\beta-1}e^s ds=\tau^{\beta-1}e^\tau\sum_{n=0}^\infty(-1)^n\frac{\Gamma(1-\beta+n)}{\Gamma(1-\beta)}\tau^{-n}
\end{equation}
for $-\pi<\arg\tau<2\pi$. Substitution of Eq.\:\eqref{A1expand} into Eq.\:\eqref{A2definition} yields
\begin{align}\label{A2Asymptotic}
    &A_2(\tau)=\sum_{n=0}^\infty\beta_n^{(1)}\tau^{-\frac{n}{2}}\nonumber\\
    &+T_1\Theta_1(\tau)\sum_{r=0}^\infty\Delta_r^{(2)}e^\tau\sum_{n=0}^\infty\frac{d^n}{d\tau^n}(\tau^{-\frac{r}{2}-1+2\mu_2}e^{4\mu_1\tau^{\frac{1}{2}}}),
\end{align}
where $\Delta_r^{(2)}\equiv\sum_{p+q=r}B_{p+1}^{(2)}\beta_q^{(2)}$, $\beta_n^{(1)}\equiv-\frac{2}{n}\sum_{p+q=n-1}B_{p+1}^{(2)}\alpha_{q+1}^{(1)}$. Comparing Eq.\:\eqref{A2expand} and \eqref{A2Asymptotic}, we obtain
\begin{align}
    &\sum_{k=0}^\infty\alpha_k^{(2)}\tau^{-\frac{k}{2}}=\sum_{r=0}^\infty\Delta_r^{(2)}\sum_{n=0}^\infty\sum_{m=0}^n\binom{n}{m}\frac{\Gamma(2\mu_2-\frac{r}{2})\tau^{-n+m-1-\frac{r}{2}}}{\Gamma(2\mu_2-\frac{r}{2}-n+m)}\cdot\nonumber\\
    &\sum_{p=1}^m\sum_{q=0}^p(-1)^{p-q}\frac{(4\mu_1)^p}{q!(p-q)!}\frac{\Gamma(\frac{q}{2}+1)}{\Gamma(\frac{q}{2}+1-m)}\tau^{\frac{p}{2}-m}.
\end{align}
Hence, we obtain $\alpha_1^{(2)}=\alpha_2^{(2)}=0$ and
\begin{equation}
\alpha_{k+2}^{(2)}=\sum_{r+s=k}\Delta_r^{(2)}\sum_{n=\frac{s}{2}}^s\sum_{m=2n-s}^n\binom{n}{m}\frac{\Gamma(2\mu_2-\frac{r}{2})d_{(2n-s)m}(4\mu_1)^{2n-s}}{\Gamma(2\mu_2-\frac{r}{2}-n+m)},
\end{equation}
where
\begin{equation}
d_{pm}\equiv\sum_{q=0}^p\frac{(-1)^{p-q}}{q!(p-q)!}\frac{\Gamma(\frac{q}{2}+1)}{\Gamma(\frac{q}{2}+1-m)},\:\:(\mbox{for}\:\:p\geq 1).
\end{equation}

\section{Error Estimation for neglecting $G(x)$ in Eqs.\:\eqref{XInfinity} and \eqref{XZero}}\label{C}
In this appendix, we estimate the error on the wave amplitude $W_1(x)$ when $G(x)$ is neglected in Eqs.\:\eqref{XInfinity} and \eqref{XZero}. To begin with, let us denote $x^*$ as the transition point, so that
\begin{subequations}
\begin{align}
W_1(x)&\approx c_1^{\RN{1}}U(a,\xi)+c_2^{\RN{1}}V(a,\xi),\:\:\:\:\:\mbox{for}\:\:x \geq x^*,\\
W_1(x)&\approx c_1^{\RN{2}}M_{\kappa,\mu}(\xi)+c_2^{\RN{2}}M_{\kappa,-\mu}(\xi),\:\:\mbox{for}\:\:x\leq x^*,
\end{align}
\end{subequations}
where $a\equiv \frac{3}{4}$ and $\xi\equiv \sqrt{\frac{i\gamma}{2}}(x+\frac{\alpha}{\gamma})$ for $x \geq x^*$, and $\kappa\equiv (\frac{3i\gamma}{2}-\frac{\alpha^2}{4})^{-\frac{1}{2}}(\frac{|f|^2}{4}-\frac{i\alpha}{8})$, $\mu\equiv \frac{1}{4}$ and $\xi\equiv \sqrt{\frac{3i\gamma}{2}-\frac{\alpha^2}{4}} x$ for $x\leq x^*$. For $x\geq x^*$, the original equation which governs $W_1$ becomes
\begin{subequations}
\begin{gather}
    \frac{d^2W_1}{d\xi^2}-\left(\frac{\xi^2}{4}+a\right)W_1=\frac{2i}{\gamma}G^{\RN{1}}(\xi)W_1,\\
    G^{\RN{1}}(\xi)\equiv\frac{g_1^{\RN{1}}}{\sqrt{\frac{2}{i\gamma}}\xi-\frac{\alpha}{\gamma}}+\frac{g_2^{\RN{1}}}{\left(\sqrt{\frac{2}{i\gamma}}\xi-\frac{\alpha}{\gamma}\right)^2},
\end{gather}
\end{subequations}
where $g_1^{\RN{1}}\equiv \frac{|f|^2}{4}-\frac{i\alpha}{8}$ and $g_2^{\RN{1}}\equiv\frac{3}{16}$. To solve this inhomogeneous differential equation, we apply the method of variation of parameters. Denoting $\overline{W}_1\equiv c_1^{\RN{1}}U(a,\xi)+c_2^{\RN{1}}V(a,\xi)$, we obtain a Volterra integral equation
\begin{equation}\label{Volterra}
    W_1(\xi)=\overline{W}_1(\xi)+\int_{\sqrt{i}\infty}^\xi K^{\RN{1}}(\xi,\zeta)G^{\RN{1}}(\zeta)W_1(\zeta)d\zeta,
\end{equation}
where $K^{\RN{1}}(\xi,\zeta)\equiv \frac{\sqrt{2\pi}i}{\gamma}[-U(a,\xi)V(a,\zeta)+V(a,\xi)U(a,\zeta)]$ is the kernel for integral transform. Eq.\:\eqref{Volterra} can also be written as
\begin{equation}\label{RealVolterra}
     \mathcal{W}_1(x)=\overline{\mathcal{W}}_1(x)+\int_{\infty}^x\mathcal{K}^{\RN{1}}(x,y)G^{\RN{1}}(y)\mathcal{W}_1(y)dy,
\end{equation}
where $\mathcal{W}_1(x)\equiv W_1(\xi(x))$, $\mathcal{K}^{\RN{1}}(x,y)\equiv \sqrt{\frac{i\gamma}{2}}K^{\RN{1}}(\xi(x),\zeta(y))$, and $G^{\RN{1}}(x)=\sum_{n=1}^2g^{(\RN{1})}_nx^{-n}$. If we define an error integral $\mathcal{E}^{\RN{1}}(x)\equiv\int_{x}^\infty|\mathcal{K}^{\RN{1}}(x,y)G^{\RN{1}}(y)|dy$, and denote $M^{\RN{1}}(x)\equiv \sup_{y\in(x,\infty)}|\overline{\mathcal{W}}_1(y)|$, we obtain
\begin{align}
    &\mathcal{W}_1(x)=\overline{\mathcal{W}}_1(x)+\int_{\infty}^x \mathcal{K}^{\RN{1}}(x,y)G^{\RN{1}}(y)\overline{\mathcal{W}}_1(y)dy+\nonumber\\
    &\int_{\infty}^x \mathcal{K}^{\RN{1}}(x,y)G^{\RN{1}}(y)\int_{\infty}^{y} \mathcal{K}^{\RN{1}}(x,y^\prime)G^{\RN{1}}(y^\prime)W_1(y^\prime)dy^\prime dy+\cdots,
\end{align}
which yields an estimation of error on $\mathcal{W}_1(x)$
\begin{equation}\label{ErrorRegion1}
\varepsilon^{\RN{1}}(x)\equiv |\mathcal{W}_1(x)-\overline{\mathcal{W}}_1(x)|\leq M^{\RN{1}}(x)(e^{\mathcal{E}^{\RN{1}}(x)}-1).
\end{equation}
Similarly, for $x\leq x^*$, the original equation which governs $W_1$ can be written as
\begin{subequations}
\begin{gather}\label{WhittakerEquation}
    \frac{d^2W_1}{d\xi^2}+\left(-\frac{1}{4}+\frac{\kappa}{\xi}+\frac{\frac{1}{4}-\mu^2}{\xi^2}\right)W_1=-G^{\RN{2}}(\xi)W_1,\\
    G^{\RN{2}}(\xi)\equiv \frac{\alpha\gamma}{8}\frac{\xi}{\lambda^3}+\frac{\gamma^2}{16}\frac{\xi^2}{\lambda^4}\:\:\mbox{and}\:\:\lambda\equiv\sqrt{\frac{3i\gamma}{2}-\frac{\alpha^2}{4}}. 
\end{gather}
\end{subequations}
The inhomogeneous differential equation Eq.\:\eqref{WhittakerEquation} is equivalent to the following Volterra integral equation
\begin{equation}\label{VolterraTwo}
    W_1(\xi)=\overline{W}_1(\xi)+\int_0^\xi K^{\RN{2}}(\xi,\zeta)G^{\RN{2}}(\zeta)W_1(\zeta)d\zeta,
\end{equation}
where $K^{\RN{2}}(\xi,\zeta)\equiv 2(-M_{\kappa,\mu}(\xi)M_{\kappa,-\mu}(\zeta)+M_{\kappa,-\mu}(\xi)M_{\kappa,\mu}(\zeta))$ is the kernel for the integral transform, and $\overline{W}_1(\xi)\equiv c_1^{\RN{2}}M_{\kappa,\mu}(\xi)+c_2^{\RN{2}}M_{\kappa,-\mu}(\xi)$. Eq.\:\eqref{VolterraTwo} can also be written as
\begin{equation}
 \mathcal{W}_1(x)=\overline{\mathcal{W}}_1(x)+\int_0^x\mathcal{K}^{\RN{2}}(x,y)G^{\RN{2}}(y)\mathcal{W}_1(y)dy,
\end{equation}
where $\mathcal{W}_1(x)\equiv W_1(\lambda x)$, $G^{\RN{2}}(x)\equiv \sum_{n=1}^2 g_n^{\RN{2}}x^n
$, $g_1^{\RN{2}}\equiv \frac{\alpha\gamma}{8\lambda^2}$, $g_2^{\RN{2}}\equiv \frac{\gamma^2}{16\lambda^2}$, and $\mathcal{K}^{\RN{2}}(x,y)\equiv\lambda K^{\RN{2}}(\lambda x,\lambda y)$. Following the method used in deriving Eq.\:\eqref{ErrorRegion1}, we may denote $M^{\RN{2}}(x)\equiv \sup_{y\in(0,x)}|\overline{\mathcal{W}}_1(y)|$ and $\mathcal{E}^{\RN{2}}(x)\equiv\int_0^x|\mathcal{K}^{\RN{2}}(x,y)G^{\RN{2}}(y)|dy$, which yields an estimation of error on $\mathcal{W}_1(x)$
\begin{equation}
    \varepsilon^{\RN{2}}(x)\equiv|\mathcal{W}_1(x)-\overline{\mathcal{W}}_1(x)|\leq M^{\RN{2}}(x)(e^{\mathcal{E}^{\RN{2}}(x)}-1).
\end{equation}
Hence, the transition point $x^*$ may be determined by the condition $\varepsilon^{\RN{1}}(x^*)=\varepsilon^{\RN{2}}(x^*)$, or explicitly by $M^{\RN{1}}(x^*)(e^{\mathcal{E}^{\RN{1}}(x^*)}-1)=M^{\RN{2}}(x^*)(e^{\mathcal{E}^{\RN{2}}(x^*)}-1)$. 

We now estimate the error integral $\mathcal{E}^{\RN{1}}(x)$ for $x\rightarrow\infty$. For $|\xi|\rightarrow\infty$ with $\arg\xi=\frac{\pi}{4}$, the parabolic cylinder functions $U(a,\xi)$ and $V(a,\xi)$ have the following asymptotic expressions
\begin{align}
U(a,\xi)&\approx e^{-\frac{\xi^2}{4}}\xi^{-a-\frac{1}{2}},\nonumber\\
V(a,\xi)&\approx\sqrt{\frac{2}{\pi}}e^{\frac{\xi^2}{4}}\xi^{a-\frac{1}{2}}+\frac{i}{\Gamma(\frac{1}{2}-a)}e^{-\frac{\xi^2}{4}}\xi^{-a-\frac{1}{2}},
\end{align}
where $\xi^2\approx \frac{i\gamma}{2}x^2$. Hence, a direct computation yields
\begin{align}\label{IntegralEstimation}
    &-U(a,\xi(x))V(a,\zeta(y))+V(a,\xi(x))U(a,\zeta(y))\nonumber\\
    \approx& -(\sqrt{\frac{i\gamma}{2}}x)^{-a-\frac{1}{2}}[\sqrt{\frac{2}{\pi}}e^{\frac{-i\gamma(x^2-y^2)}{8}}(\sqrt{\frac{i\gamma}{2}}y)^{a-\frac{1}{2}}\nonumber\\
    &+\frac{ie^{\frac{-i\gamma (x^2+y^2)}{8}}}{\Gamma(\frac{1}{2}-a)}(\sqrt{\frac{i\gamma}{2}}y)^{-a-\frac{1}{2}}]+[\sqrt{\frac{2}{\pi}}e^{\frac{i\gamma (x^2-y^2)}{8}}(\sqrt{\frac{i\gamma}{2}}x)^{a-\frac{1}{2}}\nonumber\\
    &+\frac{ie^{\frac{-i\gamma (x^2+y^2)}{8}}}{\Gamma(\frac{1}{2}-a)}(\sqrt{\frac{i\gamma}{2}}x)^{-a-\frac{1}{2}}](\sqrt{\frac{i\gamma}{2}}y)^{-a-\frac{1}{2}}.
\end{align}
In our case, $a=\frac{3}{4}$, and thus we have $x^a\gg x^{-a}$ for large $x$. Hence, the error integral $\mathcal{E}^{\RN{1}}(x)$ for $x\rightarrow\infty$ is given by
\begin{align}\label{ErrorIntegralForLargeX}
    \mathcal{E}^{\RN{1}}(x)&\approx\frac{2}{\gamma}\int \sqrt{[(\frac{y}{x})^a-(\frac{x}{y})^a]^2+4\sin^2\frac{\gamma}{8}(y^2-x^2)}(xy)^{-\frac{1}{2}}d\mu\nonumber\\
    &\approx \frac{2}{\gamma}\int[(\frac{y}{x})^a-(\frac{x}{y})^a](xy)^{-\frac{1}{2}}d\mu\nonumber\\
    &\approx\frac{2}{\gamma}\sum_{n=1}^2|g^{\RN{1}}_n|(x^{-a-\frac{1}{2}}\int_x^\infty y^{a-\frac{1}{2}-n}dy-x^{a-\frac{1}{2}}\int_x^\infty y^{-a-\frac{1}{2}-n}dy),
\end{align}
where $d\mu\equiv \left|G^{\RN{1}}(y)\right|dy$. Hence, the error integral $\mathcal{E}^{\RN{1}}(x)$ diverges, as the integral $\int_x^\infty y^{a-\frac{1}{2}-n}dy$ in Eq.\:\eqref{ErrorIntegralForLargeX} diverges for $a=\frac{3}{4}$ and $n=1$. To fix this problem, one may modify the kernel for integral transform in region I. From the initial condition $\overline{\mathcal{W}}_1(\infty)\equiv c_1^{\RN{1}}U(a,\xi(\infty))+c_2^{\RN{1}}V(a,\xi(\infty))=0$, one obtains $c_2^{\RN{1}}=0$. Hence, one may use the following transformation
\begin{subequations}
\begin{align}
    \tilde{\mathcal{K}}(x,y)&\equiv x^m\mathcal{K}(x,y)y^{-m},\\
    \tilde{\mathcal{W}}_1(x)&\equiv x^m\mathcal{W}_1(x).
\end{align}
\end{subequations}
Then, in region I, we obtain
\begin{align}\label{ModifiedErrorIntegral}
    \tilde{\mathcal{E}}^{\RN{1}}(x)&\approx\frac{2}{\gamma}\int(\frac{x}{y})^m \sqrt{[(\frac{y}{x})^a-(\frac{x}{y})^a]^2+4\sin^2\frac{\gamma}{8}(y^2-x^2)}(xy)^{-\frac{1}{2}}d\mu\nonumber\\
    &<\frac{2}{\gamma}\int(\frac{x}{y})^m\left[(\frac{y}{x})^a-(\frac{x}{y})^a+2|\sin\frac{\gamma}{8}(y^2-x^2)|\right](xy)^{-\frac{1}{2}}d\mu\nonumber\\
    &\leq \frac{2}{\gamma}\sum_{n=1}^2|g^{\RN{1}}_n|(x^{m-\frac{5}{4}}\int_x^\infty y^{\frac{1}{4}-m-n}dy-x^{m+\frac{1}{4}}\int_x^\infty y^{-\frac{5}{4}-m-n}dy)\nonumber\\
    &+\frac{4}{\gamma}\sum_{n=1}^2|g^{\RN{1}}_n|x^{m-\frac{1}{2}}\int_x^\infty y^{-\frac{1}{2}-m-n}dy.
\end{align}
For $m>\frac{1}{4}$, the two integrals in the second last lines in Eq.\:\eqref{ModifiedErrorIntegral} becomes
\begin{align}
&x^{m-\frac{5}{4}}\int_x^\infty y^{\frac{1}{4}-m-n}dy-x^{m+\frac{1}{4}}\int_x^\infty y^{-\frac{5}{4}-m-n}dy\nonumber\\
=&\frac{x^{-n}}{m+n-\frac{5}{4}}-\frac{x^{-n}}{m+n+\frac{1}{4}}=\frac{\frac{3}{2}x^{-n}}{(m+n-\frac{5}{4})(m+n+\frac{1}{4})}.
\end{align}
Hence, the error integral $\tilde{\mathcal{E}}^{\RN{1}}(x)$ is bounded by
\begin{align}
    \tilde{\mathcal{E}}^{\RN{1}}(x)&\leq \frac{3}{\gamma}\sum_{n=1}^2\frac{|g^{\RN{1}}_n|x^{-n}}{(m+n-\frac{5}{4})(m+n+\frac{1}{4})}+\frac{4}{\gamma}\sum_{n=1}^2\frac{|g^{\RN{1}}_n|x^{-n}}{m+n-\frac{1}{2}}\nonumber\\
    &<\frac{1}{\gamma}\left[\frac{3}{(m-\frac{1}{4})(m+\frac{5}{4})}+\frac{4}{m+\frac{1}{2}}\right]|G^{\RN{1}}(x)|.
\end{align}
Clearly, the error integral $\tilde{\mathcal{E}}^{\RN{1}}(x)$ vanishes for $x\rightarrow\infty$, and thus $\tilde{\mathcal{E}}^{\RN{1}}(x)$ is bounded on $(x,\infty)$. Moreover, for $m<\frac{5}{4}$, we have $\lim_{x\rightarrow\infty}x^m\left|U(a,\xi(x))\right|\approx (\sqrt{\frac{\gamma}{2}})^{-\frac{5}{4}}x^{m-\frac{5}{4}}\rightarrow 0$, and thus $\tilde{M}^{\RN{1}}(x)\equiv \sup_{y\in(0,x)}y^m|\overline{\mathcal{W}}_1(y)|$ is also bounded on $(x,\infty)$. To be precise, we may choose $m=a=\frac{3}{4}$, which yields
\begin{equation}
    \mathcal{E}^{\RN{1}}(x)\equiv|\mathcal{W}_1(x)-\overline{\mathcal{W}}_1(x)|\leq x^{-a}\tilde{M}^{\RN{1}}(x)(e^{\tilde{\mathcal{E}}^{\RN{1}}(x)}-1).
\end{equation}
Hence, the transition point $x^*$ may now be determined by the condition $(x^*)^{-a}\tilde{M}^{\RN{1}}(x^*)(e^{\tilde{\mathcal{E}}^{\RN{1}}(x^*)}-1)=M^{\RN{2}}(x^*)(e^{\mathcal{E}^{\RN{2}}(x^*)}-1)$.

\end{appendix}

\end{document}